\documentclass[aps,prc,nofootinbib,groupedauthor,twocolumn,10pt,floatfix]{revtex4-2}

\usepackage{graphicx}
\usepackage{amsmath}
\usepackage{hyperref}
\usepackage{isotope}

\usepackage[svgnames]{xcolor}

\usepackage{liealg}  \usepackage{revtextools}

\newcommand{\MeV}{{\mathrm{MeV}}}

\newcommand{\fm}{{\mathrm{fm}}}
\newcommand{\Nmax}{{N_\text{max}}}
\newcommand{\Nex}{{N_\text{ex}}}
\newcommand{\hw}{{\hbar\omega}}  

\begin{document}

\title{Deformation-driven intruder states in light island-of-inversion nuclei}

\author{Calvin W.~Johnson}
\affiliation{Department of Physics, San Diego State University, San Diego, California 92182-1233, USA}

\author{Mark A.~Caprio}
\affiliation{Department of Physics and Astronomy, University of Notre Dame, Notre Dame, Indiana 46556-5670, USA}

\author{Shwetha L.~Vittal}
\affiliation{Department of Physics and Astronomy, University of Notre Dame, Notre Dame, Indiana 46556-5670, USA}

\date{\today}

\begin{abstract}
Islands of inversion occur when the nuclear ground state 
is dominated by intruder configurations, specifically particle-hole 
excitations across shell gaps,
rather than by the naive spherical shell-model expectation of filled shell configurations. Using the realistic and rigorous no-core shell model, we are able to 
confirm that deformation drives these intruder states in the light halo nuclides 
$\isotope[11]{Li}$ and $\isotope[29]{F}$.
In small model spaces, these deformed intruders lie high in energy with respect
to spherical normal states; as the model space size increases, the intruders
energetically approach, albeit slowly, the normal states.   This provides further strong evidence of the
connection between shape deformation/coexistence and islands of inversion, as well demonstrating as the computational challenges in rigorously modeling this phenomenon. Our
results also suggest halo states can be strongly deformed and/or strongly mixed
with normal spherical states.
\end{abstract}

\maketitle

\section{Introduction}

Simple models can be challenged by messy realities; adapting to reality can lead
to more insightful models. As an early example of this, when the spherical
independent-particle shell-model of nuclei was confronted by collective
quadrupole deformation and dynamics, this led first to the deformed Nilsson
model~\cite{nilsson1955:model-NUMBER-AS-PAGE,lawson1980:shell}, and then to
deformation in Elliott's $\grpsu{3}$
picture~\cite{elliott1958:su3-part1,elliott1958:su3-part2,elliott1963:su3-part3,elliott1968:su3-part4,harvey1968:su3-shell,vanisacker2011:elliott-legacy}.
While such models of deformation were initially applied within the traditional
valence space of the shell model, that is, involving only excitations of
nucleons within a major shell, deformation (indeed, much larger deformation) can
also be associated with intruder states.  These are states that, in a
single-particle picture, ought to be high in energy, since they involve
particle-hole excitations across a shell gap, but which, due to collectivity,
are dramatically lowered in energy, and thus ``intrude'' into the low-energy
excitation spectrum.

If the collectivity driving the intruder state is quadrupole in nature, such a
deformation-driven intruder naturally connects to the idea of shape coexistence:
spherical (or less deformed) normal states and deformed intruder states coexist
in the low-lying excitation spectrum, therefore potentially mixing as well.
Intruder states are invoked, in this way, to explain regions of shape
coexistence across the nuclear chart (see
Ref.~\cite{heyde2011:shape-coexistence}, especially their Figs.~2 and 8).  In
particular, intruders (and shape coexistence) involving one-particle, one-hole
or two-particle, two-hole excitations across a major shell gap proliferate near
shell closures.  Other examples of intruders include four-particle, four-hole
intruders arising near the $N=Z$ line, such as the Hoyle state in
$\isotope[12]{C}$~\cite{freer2014:12c-hoyle,PhysRevLett.106.192501}, the
superdeformed first excited $0^+$ state at $6.05\,\MeV$ in
$\isotope[16]{O}$~\cite{PhysRevLett.65.1325,warburton1992:16o-shell-4ho}, and $0^+$
states at $3.35$ and $5.21\,\MeV$ in
$\isotope[40]{Ca}$~\cite{PhysRevC.75.054317}.  These may also be deformation
driven.

When intruders descend far enough in energy to strongly mix with, or even
become, the ground state, we land on the notorious islands of
inversion~\cite{PhysRevC.41.1147,PhysRevC.90.014302}, which occur near shell
closures.  The first identified example of inversion was
$\isotope[11]{Be}$~\cite{PhysRevLett.4.469}, for which the ground state has
$J^\pi = 1/2^+$, whereas the naive independent-particle shell model picture
predicts $1/2^-$.  Here one-particle, one-hole excitations across the shell gap
at $N=8$ yield a so-called intruder state of non-normal parity for the ground
state of $\isotope[11]{Be}$ (in what is sometimes termed parity inversion).
While it was initially proposed that inversion in $\isotope[11]{Be}$ arose simply from
shifts of spherical single-particle energies~\cite{PhysRevLett.4.469}, later it
was argued that the intruder arises from the extra binding energy the nucleus
can obtain by collectively transitioning to a deformed shape, even if that
transition involves excursions beyond the valence space.  Such a scenario is
easily perceived using the Nilsson model (see Sec.~5-3 of
Ref.~\cite{bohr1998:v2}, as well as
Refs.~\cite{hamamoto2007:11be-12be-nilsson,macchiavelli2018:11be-12be-nilsson}).
Alternatively, as in neighboring $\isotope[12]{Be}$, two-particle, two-hole
excitations can yield intruder ground states of normal parity.

While deformation has been widely accepted as a qualitative explanation of the
islands of
inversion~\cite{RevModPhys.77.427,heyde2011:shape-coexistence,otsuka2020:shell-structure,nowacki2021:neutron-rich},
most previous work has focused on the particle-hole
structure~\cite{PhysRevC.41.1147,PhysRevC.90.014302}.  Our goal here is to
quantitatively account for the deformation in a way that is realistic and
rigorous, namely through the use of no-core configuration-interaction,
also known as no-core shell model (NCSM)~\cite{navratil2000:12c-ncsm}, calculations.  We focus on the
light halo nuclei $\isotope[11]{Li}$ and $\isotope[29]{F}$, 
which are known (or thought) to have two-particle, two-hole intruder components to their ground states,
and trace the evolution of deformed
intruder states versus normal spherical states as the model space increases.
Specifically, to illuminate deformation we carry out group-theoretical
decompositions of the wave functions, similar to work on ground state inversion
in $\isotope[12]{Be}$~\cite{mccoy2024:12be-shape}, as well as computing the quadrupole
moment of the neutron density.

A natural mechanism by which deformation (and rotation) arises within a
spherical harmonic oscillator shell model picture, and which we shall see plays
an important role for intruders, is provided by Elliott's $\grpsu{3}$
symmetry~\cite{elliott1958:su3-part1,elliott1958:su3-part2,elliott1963:su3-part3,elliott1968:su3-part4,harvey1968:su3-shell}.
This symmetry reorganizes the shell model space into irreducble representations
(irreps) of $\grpsu{3}$.  All states within the same irrep have (at least in
some appropriately defined sense) the same quadrupole deformation, which can be
deduced from the $\grpsu{3}$ quantum numbers
$(\lambda,\mu)$~\cite{castanos1988:su3-shape}.  A quadrupole-quadrupole
internucleon interaction energetically favors the most quadrupole-deformed
$\grpsu{3}$ irrep, known as the ``leading'' irrep~\cite{harvey1968:su3-shell},
within the shell model space.  If we now make the traditional division of the
shell model space into ``$\Nex \hw$'' subspaces ($0\hw$, $1\hw$, $2\hw$,
\textit{etc.}), based on the number $\Nex$ of harmonic oscillator excitations
relative to the lowest Pauli-allowed filling of the oscillator
shells,\footnote{Note that the $0\hw$ space is the ordinary zero-particle,
zero-hole valence space, while one-particle, one-hole excitations by a single
oscillator shell constitute the $1\hw$ space, and two-particle, two-hole
excitations by a single oscillator shell lie within the $2\hw$ space.} the
Elliott picture suggests that, within each $\Nex \hw$ subspace, a
quadrupole-quadrupole interaction will again favor the leading irrep
\textit{within that subspace}.  This naturally leads to the
proposition~\cite{rowe2006:coexistence-shell-u3,dreyfuss2013:12c-sp-rotation,rowe2020:shape-coexistence-algebraic,nowacki2021:neutron-rich}
that deformed intruder states arise when the highly collective, deformed states
in the leading $\grpsu{3}$ irrep of a given $\Nex \hw$ ``excited'' space, are
sufficiently lowered in energy so as to compensate for the penalty of
$\approx\Nex \hw$ in single-particle energy incurred by accessing that space
(see also Ref.~\cite{nowacki2021:neutron-rich} for variations on this basic
theme).

The (no-core) shell-model basis states  we use in our calculations are
antisymmetrized products (Slater determinants) of harmonic oscillator
single-particle states.  Therefore, they may again be characterized by a number
$\Nex$ of oscillator excitations, relative to the lowest filling of oscillator
shells.  A given, truncated NCSM model space is then labeled by
$\Nmax$, the maximum number of excitations allowed. The full
many-body space is in principle recovered in the limit $\Nmax\rightarrow\infty$.
(The underlying oscillator single-particle basis is also characterized by a
radial length scale, or, equivalently, oscillator energy parameter $\hw$.  As
$\Nmax\rightarrow\infty$, results should ``converge'' to values independent of
both $\hw$ and $\Nmax$.)

While there is formally no valence space in the NCSM, it is nonetheless natural
to identify the $\Nmax=0$ or $0\hbar\omega$ space with the naive shell model
valence space.  Then we may classify as ``normal'' those states which are
dominated by configurations in the valence space, and as ``intruders'' those
dominated by configurations outside the valence space.

Since the earliest days of the NCSM, intruder states have been recognized among
the results of NCSM
calculations~\cite{PhysRevC.64.051301,PhysRevC.66.024314,navratil2003:ncsm-3n,forssen2005:ncsm-9be-11be}.
However, these states are plagued by slow convergence with increasing model
space.  That is, they start out far higher in energy, in truncated calculations
at computationally feasible $\Nmax$, than they would lie if calculated in the
full many-body space, for a given internucleon interaction.  They then descend
in energy only slowly with increasing $\Nmax$.  Perhaps in part for this reason,
intruder states have also subsequently largely been neglected in NCSM
calculations. Nonetheless, internucleon interactions which have been
``softened'' through the similarity renormalization group
(SRG)~\cite{PhysRevC.75.061001,PhysRevC.83.034301} help to tame the convergence
behavior of NCSM calculations, including for intruder states.

In this context,
intruder states in the NCSM have more recently been revisited.  They have been
found to form highly deformed rotational band structures, with much larger
moment of inertia than the normal
states~\cite{caprio2019:bebands-sdanca19,caprio2020:bebands}. They have also
been interpreted in terms of their Elliott's $\grpsu{3}$, as well as
symplectic $\grpsptr$~\cite{rowe1996:sp3r-dynamical-symmetry}, symmetry
content~\cite{mccoy2018:diss,mccoy2020:spfamilies,dytrych2020:emergent-symmetry,zbikowski2021:beyond-elliott,caprio2022:10be-shape-sdanca21,PhysRevLett.128.202503,mccoy2024:12be-shape}.
The results for various light nuclei~--- including
$\isotope[7]{Be}$~\cite{mccoy2018:diss,mccoy2020:spfamilies},
$\isotope[8]{Be}$~\cite{mccoy2018:diss,PhysRevLett.128.202503},
$\isotope[10]{Be}$~\cite{caprio2022:10be-shape-sdanca21}, and
$\isotope[12]{Be}$~\cite{mccoy2024:12be-shape}~--- support the proposition that
$2\hw$ intruder states stem from the leading $\grpsu{3}$ irrep in the $2\hw$
space.

\begin{figure}
  \includegraphics[width=.90\hsize]{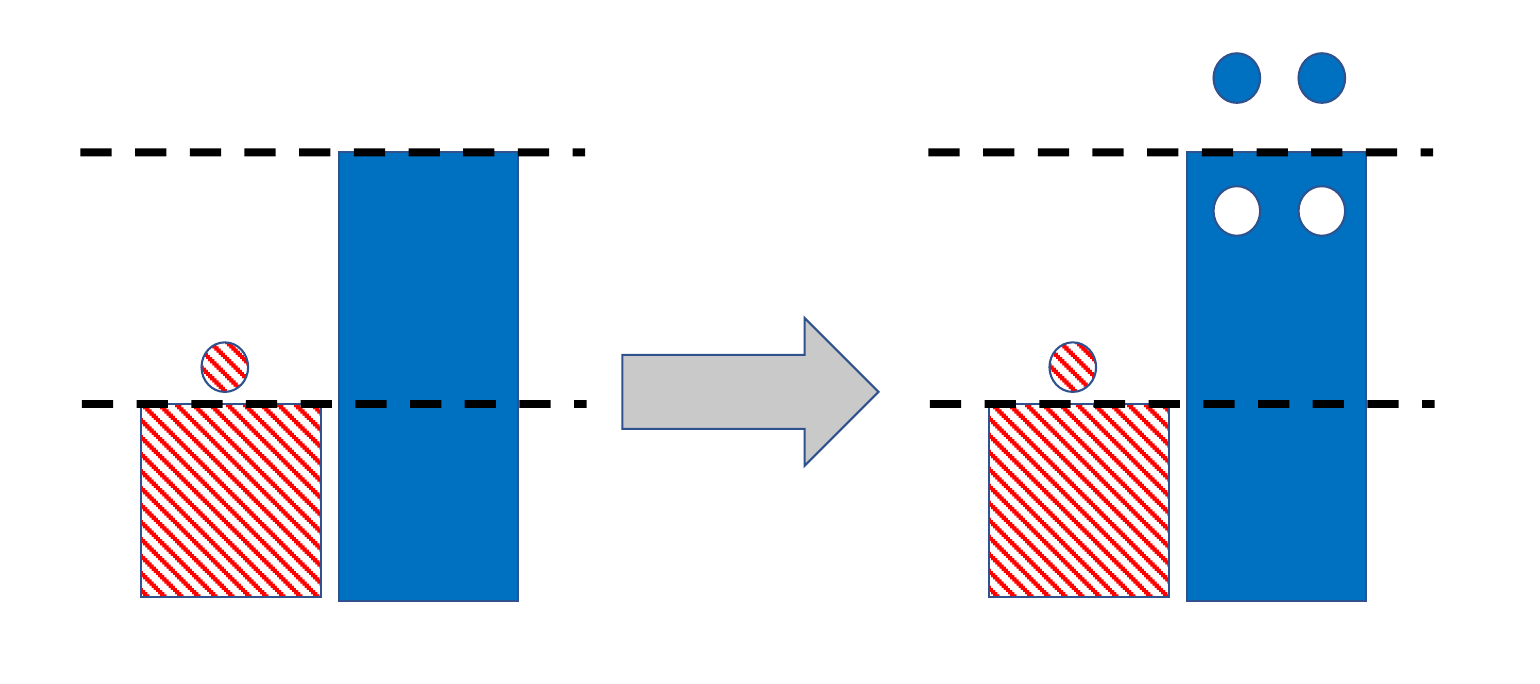}
  \caption{Schematic of normal (left) versus two-particle, two-hole
    intruder (right) configurations. Dashed lines represent spherical shell
    closures.  For our example of $\isotope[11]{Li}$, the filled proton shell (hashed
    fill) is $0s$ while the filled neutron shells (solid fill) consist of
    $0s$-$0p$. For $\isotope[29]{F}$, the filled proton and neutron shells are
    $0s$-$0p$ and $0s$-$0p$-$1s0d$, respectively. }
  \label{fig:li11schematic}
\end{figure}

In the present work, we turn to light island-of-inversion nuclei at $A=11$ and
$29$.  We aim to understand and highlight, through \textit{ab initio} NCSM
calculations, the role of deformation-driven intruders in the structure of these
nuclei.  The $N=8$ nucleus $\isotope[11]{Li}$ provides one of the earliest
examples of a neutron halo ground
state~\cite{tanihata1985:radii-11li-halo,tanihata1992:11li-halo-density-correlation}.  Neutron
knockout experiments show nearly equal contributions of neutron $1s_{1/2}$ and
$0p_{1/2}$ orbitals~\cite{PhysRevLett.83.496}, providing dramatic evidence that
the ground state arises from mixing of normal and intruder states. Stepping up 
by a major shell, one notices  $\isotope[29]{F}$ shares many parallels with 
$\isotope[11]{Li}$: in the naive spherical shell model, one would expect the
lowest configuration to consist of filled neutron oscillator shells ($0s$-$0p$ in the case of $\isotope[11]{Li}$, and $0s$-$0p$-$1s0d$ for
$\isotope[29]{F}$) and a single proton outside filled shells ($0s$ for
$\isotope[11]{Li}$, and $0s$-$0p$ for $\isotope[29]{F}$).  
Fig.~\ref{fig:li11schematic} schematically illustrates this generic situation.
Furthermore, $\isotope[29]{F}$ is another likely halo nucleus~\cite{PhysRevLett.124.222504}, lies on or at least
  adjacent to an island of
  inversion~\cite{revel2020:28f-removal,wang2023intruder}, and has been suggested
  to have low-lying intruder structure~\cite{PhysRevC.90.014302,macchiavelli2017:29f-nilsson-rotation-aligned,PhysRevC.104.014307} or possibly
  inversion~\cite{fortunato202029f} itself.
We thus use the NCSM to illuminate
deformation-driven intruder structure in $\isotope[29]{F}$ as well.

In our NCSM calculations for $\isotope[11]{Li}$ and $\isotope[29]{F}$ 
(with the codes \texttt{BIGSTICK}~\cite{johnson2018:bigstick} and
\texttt{MFDn}~\cite{maris2010:ncsm-mfdn-iccs10,shao2018:ncci-preconditioned,code-mfdn-transitions}),
we use two sets of interaction matrix elements, to explore and illustrate the
importance of a sufficiently soft internucleon interaction.  Both are ultimately
derived from the Entem-Machleidt next-to-next-to-next-to-leading-order (N3LO)
chiral effective theory interaction~\cite{PhysRevC.68.041001}, and then softened
by SRG.  The first set of matrix elements, referred to simply as
N3LO in the following, are evolved to a resolution parameter value of $\lambda = 2.0\,\fm^{-1}$
(we do not include three-body forces).  The second set of interaction matrix 
elements are those of the Daejeon16~\cite{shirokov2016:nn-daejeon16} interaction, which
are now more drastically
softened by SRG evolution, to $\lambda = 1.5\,\fm^{-1}$, after which
phase-shift-equivalent transformations are applied (to improve the description of
binding energies and selected excitation energies in several nuclei with
$A\leq16$, without including three-body forces).

To gain insight into the structure of the wave functions thus obtained, we
decompose them into $\Nex\hw$ subspaces and then into irreps of Elliott's
$\grpsu{3}$, via the Lanczos
algorithm~\cite{whitehead1980:lanczos,gueorguiev2000:fp-su3-breaking,PhysRevC.91.034313,caprio2022:10be-shape-sdanca21}
(see
Refs.~\cite{PhysRevC.91.034313,caprio2022:10be-shape-sdanca21,mccoy2024:12be-shape}
for details of the method).  These $\grpsu{3}$ irreps can be approximately
interpreted in terms of quadrupole deformation parameters $\beta$ and
$\gamma$~\cite{castanos1988:su3-shape} for the nucleus.  While the $\grpsu{3}$
decomposition provides the most detailed insight into deformation, we also
consider how the deformation is reflected in more traditional quadrupole
observables, namely, quadrupole moments~--- both those for the proton density
distribution within the nucleus, \textit{i.e.}, the traditional electromagnetic
$E2$ moment, and those defined analogously for the neutron density
distribution.  We first consider $\isotope[11]{Li}$ in detail
(Sec.~\ref{sec:11li}), then provide an overview of the analogous results (but at
lower $\Nmax$) for $\isotope[29]{F}$ (Sec.~\ref{sec:29f}).

\begin{figure*}
\begin{minipage}{\ifproofpre{0.85}{1}\hsize}
\includegraphics[width=0.68\hsize]{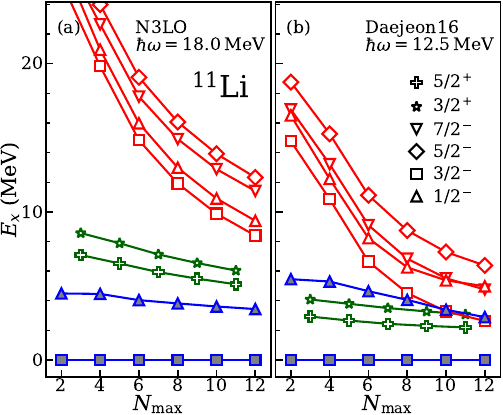}
    \hfill
    \includegraphics[width=0.30\hsize]{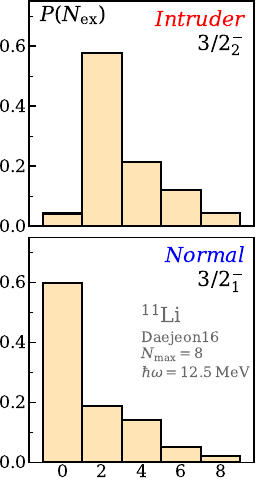}
  \end{minipage}
  \caption{
(Left)~Relative energies for the normal $1/2^-$ and $3/2^-$ (blue, shaded
    symbols), first intruder $1/2^-$ through $7/2^-$ (red, open symbols), and
    nonnormal parity first $3/2^+$ and $5/2^+$ (green, open symbols) levels of
    $\isotope[11]{Li}$, for (a)~a chiral N3LO interaction and (b)~the Daejeon16
    interaction, shown as functions of $\Nmax$ (at fixed $\hw$, as indicated).
    Although states are designated as normal or intruder in this figure
    according to what might naively be expected for the level, given the energy
    evolution, the first two $1/2^-$ levels undergo an avoided crossing
    at higher $\Nmax$ in the Daejeon16 calculations (see text).  Nonnormal
    parity excitation energies are taken relative to the normal parity ground
    state energy obtained at the next lower even value of $\Nmax$.
(Right)~Decompositions by $\Nex$ for the normal~(bottom) and intruder~(top)
    $3/2^-$ levels of $\isotope[11]{Li}$, for the Daejeon16 interaction, from
    calculations with $\Nmax=8$ and $\hw=12.5\,\MeV$.
  }
  \label{fig:li11ex}
\end{figure*}
\begin{figure*}
  \includegraphics[width=\ifproofpre{0.75}{0.8}\hsize]{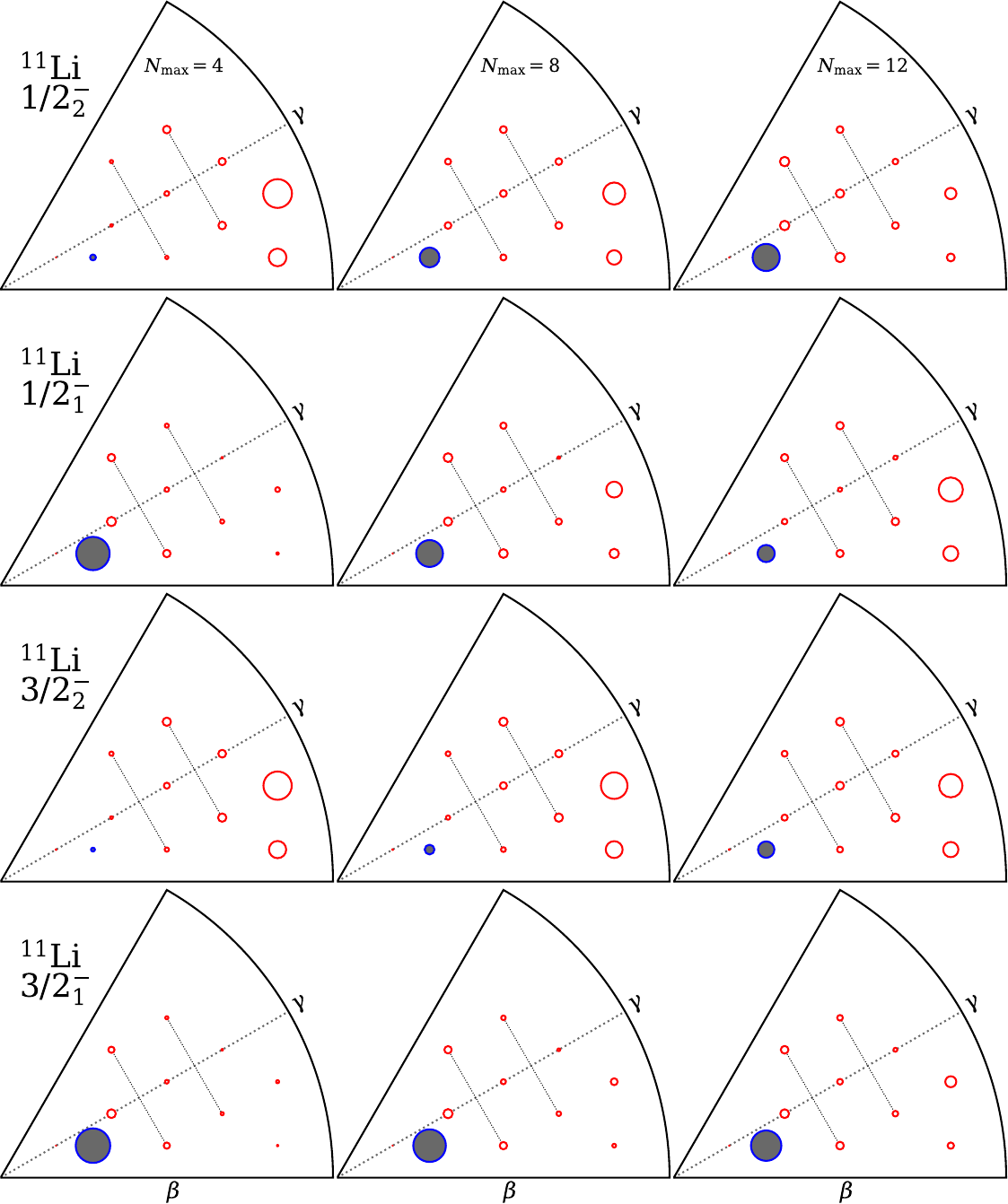}
  \caption{Decompositions by $\grpsu{3}$ jointly with $\Nex$ for the first two
    $3/2^-$ levels (bottom two rows) and first two $1/2^-$ levels (top two rows)
    of $\isotope[11]{Li}$, for the Daejeon16 interaction, in calculations with
    $\Nmax=4$~(left), $8$~(center), and $10$~(right), from calculations with
    $\hw=12.5\,\MeV$.  The $\grpsu{3}$ decompositions are shown arranged by the
    Bohr deformation variables corresponding to the given $\grpsu{3}$ quantum
    numbers, and include contributions from the $0\hw$ (blue, shaded circles)
    and $2\hw$ (red, open circles) spaces.  Contributions from irreps which are
    degenerate with respect to the $\grpsu{3}$ Casimir operator (connected by
    dotted lines) cannot be distinguished in the present decomposition (and, for
    plotting purposes, such contributions have, arbitrarily, been distributed
    equally between these irreps).  }
  \label{fig:li11decomp-wedge-normal}
\end{figure*}
\begin{figure}
  \includegraphics[width=\ifproofpre{0.60}{0.35}\hsize]{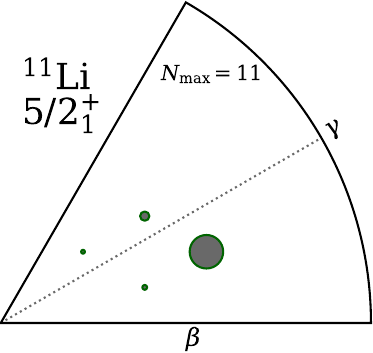}
  \caption{Decomposition by $\grpsu{3}$ for the first $5/2^+$ level of
    $\isotope[11]{Li}$, for the Daejeon16 interaction, from a calculation with
    $\Nmax=11$ and $\hw=12.5\,\MeV$.  The $\grpsu{3}$ decomposition is shown
    arranged by the Bohr deformation variables corresponding to the given
    $\grpsu{3}$ quantum numbers, and including only contributions from the $1\hw$
    space.}
  \label{fig:li11decomp-wedge-nonnormal}
\end{figure}
\begin{figure}
  \includegraphics[width=1.0\hsize]{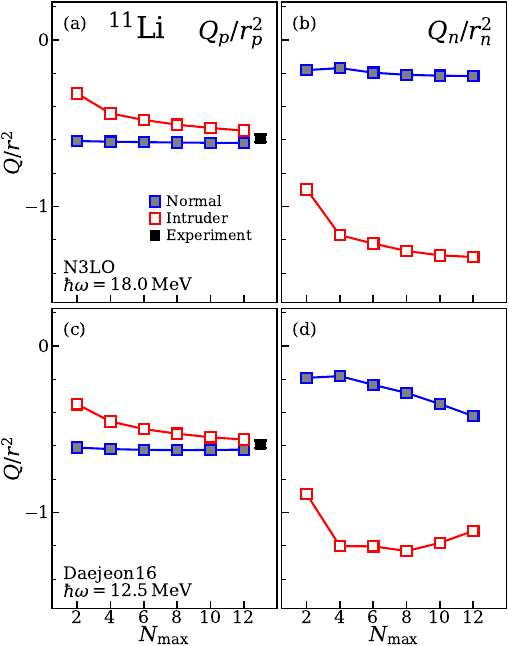}
  \caption{Dimensionless ratio $Q/r^2$ for normal (blue, shaded symbols) and
    intruder (red, open symbols) $3/2^-$ levels of $\isotope[11]{Li}$, for the
    protons~(left) and neutrons~(right), calculated with the chiral N3LO~(top)
    and Daejeon16~(bottom) interactions.  Calculated values are shown as
    functions of $\Nmax$, at fixed $\hw$ (as indicated).  The experimental value
    for $Q_p/r_p^2$~\cite{stone2016:e2-moments,angeli2013:charge-radii} for the
    $3/2^-$ ground state is shown for comparison (solid square), with the
    point-proton radius deduced from the charge radius as detailed in
    Ref.~\cite{caprio2025:emnorm2-part1}.}
  \label{fig:li11q}
\end{figure}

\section{The $N=8$ island of inversion: $\isotope[11]{Li}$}
\label{sec:11li}

The nucleus $\isotope[11]{Li}$ provided an early and dramatic example of a halo
nucleus~\cite{tanihata1992:11li-halo-density-correlation}.  The nucleus $\isotope[11]{Li}$ is well
known to be a challenge to describe fully microscopically, i.e., in terms of
three protons and eight neutrons, for example in the
NCSM~\cite{PhysRevC.57.3119,PhysRevC.79.021303,caprio2022:emnorm}, or using
fermionic molecular dynamics (FMD)~\cite{PhysRevC.84.024307}.   The FMD
calculation addresses
mixing of $(p_{1/2})^2$ and $(s_{1/2})^2$ configurations, that is, normal and
intruder states.
The nucleus $\isotope[11]{Li}$ has also been approximated as a three-body
system, typically
$\isotope[9]{Li}+2n$~\cite{esbensen1992:11li-soft-dipole,zhukov1993:6he-11li-halo-ground-state-properties,PhysRevC.42.758,PhysRevC.50.R550,PhysRevC.59.1806,PhysRevC.65.034007,brida2006effects,betan2017cooper,PhysRevC.101.034003,PhysRevC.107.014003},  allowing  $(p_{1/2})^2$-$(s_{1/2})^2$ mixing to be investigated. 

Our calculated excitation spectrum is shown in Fig.~\ref{fig:li11ex}~(left), as
a function of $\Nmax$, for both the chiral N3LO and Daejeon16 interactions (at
fixed basis parameter $\hw=18\,\MeV$ and $12.5\,\MeV$, respectively, chosen to
approximately minimize the calculated ground state energy for each interaction).
This figure highlights the different evolution with $\Nmax$ for normal (blue,
shaded symbols) and intruder (red, open symbols) states, as well as nonnormal
parity states (green, open symbols), discussed separately below.  An
illustration of how we may identify normal and intruder states from their NCSM
wave
functions~\cite{caprio2019:bebands-sdanca19,caprio2020:bebands,mccoy2024:12be-shape}
is provided in Fig.~\ref{fig:li11ex}~(right), where we display the decomposition
of the wave functions into $\Nex\hw$ components, in particular, for the $3/2^-$
levels.

The $0\hw$ shell model picture for $\isotope[11]{Li}$ gives just two states, $3/2^-$ and
$1/2^-$, corresponding to finding the valence proton in the $p_{3/2}$
or $p_{1/2}$ orbital, respectively.  Indeed, the two lowest calculated states
(blue, shaded symbols in Fig.~\ref{fig:li11ex}) match this expectation.

From their decomposition into configurations with different $\Nex$, these states
are clearly normal, as illustrated for the $3/2^-$ ground state in
Fig.~\ref{fig:li11ex} (right, bottom).  Namely, the single largest contribution
to the probability comes from $\Nex=0$ (which, in this case, represents just a
single configuration, that is, with a $p_{3/2}$ proton and a filled neutron $0p$
shell).  However, naturally for a no-core calculation [see, \textit{e.g.},
Fig.~13(a) of Ref.~\cite{caprio2020:bebands}], this $0\hw$ configuration is
``dressed'' with a tail of smaller contributions from excited configurations.

Then, several intruder states arise at higher energy in the calculation
(red, open symbols in Fig.~\ref{fig:li11ex}), starting with closely spaced $3/2^-$
and $1/2^-$ states.  (There are many more such states, of which we only show the first
$7/2^-$ and $5/2^-$ states as representative examples.)  While these start at high
energy, for low $\Nmax$, they rapidly descend with increasing $\Nmax$.

The intruder nature of the excited $3/2^-$ and $1/2^-$ states is illustrated for
the excited $3/2^-$ state in Fig.~\ref{fig:li11ex} (right, top).  Note the
negligible $0\hw$ contribution, of a few percent, while the strongest
contribution comes from $2\hw$ configurations, which are again dressed with a
tail of more highly excited oscillator configurations [similarly to,
  \textit{e.g.}, Fig.~13(d) of Ref.~\cite{caprio2020:bebands}].

Because Daejeon16 is a ``softer'' interaction, it is perhaps unsurprising that
the intruder states found in the Daejeon16 calculation
[Fig.~\ref{fig:li11ex}(b)] descend in energy much sooner (with increasing
$\Nmax$) than those found in the N3LO calculation
[Fig.~\ref{fig:li11ex}(a)]. They also start to level off in energy, suggesting
they are approaching the values they would attain in the full, untruncated model space.

As an intruder state approaches the normal state of the same angular momentum
and parity, these states can mix.  In the Daejeon16 calculations
[Fig.~\ref{fig:li11ex}(b)], it is the $1/2^-$ normal and intruder states (upward
triangles) which approach in energy first, due to the high excitation energy of
the normal $1/2^-$ state, as well as a relatively smaller spacing between the
intruder $3/2^-$ and $1/2^-$ states.  The excited $1/2^-$ state initially tracks
the other intruder states in energy.  However, at higher $\Nmax$, it appears to
hit a floor, while the other intruder states continue to fall in energy.  On the
other hand, the normal $1/2^-$ state appears to be ``pushed down'' in excitation
energy for higher $\Nmax$.  The two levels approach to within $\approx 2
\,\MeV$, but no closer.  This behavior is suggestive of the level repulsion
arising in an avoided crossing~\cite{casten2000:ns}.  

To obtain insight into the role of deformation in the intruder states (and thus
the intruder component to the $\isotope[11]{Li}$ halo ground state) and the
prospective role of mixing as well, we now decompose the wave function into
components in irreps of Elliott's $\grpsu{3}$.  As previously noted, these
irreps may be interpreted (approximately) in terms of the standard quadrupole
shape parameters $\beta$ and $\gamma$~\cite{castanos1988:su3-shape}.  In
Fig.~\ref{fig:li11decomp-wedge-normal}, we display the structure of the wave
functions for the first two $3/2^-$ states
[Fig.~\ref{fig:li11decomp-wedge-normal} (bottom two rows)], including the ground
state, and the first two $1/2^-$ states [Fig.~\ref{fig:li11decomp-wedge-normal} (top two rows)], in calculations for
increasing $\Nmax$ from $\Nmax=4$ (left) to $12$ (right).  These are decomposed
first in terms of excitation quanta $\Nex$, and then, within each $\Nex\hw$
space, by $\grpsu{3}$ irreps $(\lambda,\mu)$.  These are arranged by their
location in the $(\beta,\gamma)$ plane, where the area of each circle is
proportional to the probability (\textit{i.e.}, contribution to the wave
function norm) from that given $\Nex$ and $(\lambda,\mu)$.  Specifically, we
show the contributions from the $0\hw$ (blue, shaded circles) and $2\hw$ (red,
open circles) spaces.

As already noted [Fig.~\ref{fig:li11ex} (right)], the $3/2^-_1$ (normal) state in
these calculations is largely $0\hw$, while the $0\hw$ contribution to the
$3/2^-_2$ (intruder) state is comparatively small.  Now, from
Fig.~\ref{fig:li11decomp-wedge-normal} (bottom two rows), we can see that the
shapes of these states differ markedly.  The normal $3/2^-_1$ state is dominated
by the unique $0\hw$ irrep $(1,0)$ of $\grpsu{3}$ (blue, shaded circle, at left
in the $\beta$-$\gamma$ plot), while the intruder $3/2_2^-$ state has its
largest contributions from highly deformed $2\hw$ $(5,2)$ and $(6,0)$ irreps
(red, open circles, at right in $\beta$-$\gamma$ plot).  The $(6,0)$ irrep
represents a prolate, axially symmetric shape (or as close to this as can be
attained in the $2\hw$ model space for $\isotope[11]{Li}$), but the $(5,2)$
irrep (which, with marginally larger deformation, is nominally the leading
irrep) lies just to the prolate side of the dotted line indicating maximal
triaxiality ($\gamma=30^\circ$).  Thus, the $\grpsu{3}$ decompositions indicate
both maximal deformation and significant deviations from axial symmetry for the
intruder state.

Moreover, with increasing $\Nmax$, we see an intruder component (namely, those same maximally
deformed $2\hw$ irreps) ``growing in'' to the ground state wave function, and,
conversely, a normal component developing in the excited $3/2^-$ state wave
function.  Thus, although level crossing behavior is not yet clearly apparent
from inspection of the energies for the $3/2^-$ states [Fig.~\ref{fig:li11ex}
  (left)], it is manifest from the wave functions.

The mixing picture is even more clearly borne out for the two $1/2^-$ states
[Fig.~\ref{fig:li11decomp-wedge-normal} (top two rows)].  At low $\Nmax$~(left),
the $1/2^-_1$ state is clearly normal, with a $\grpsu{3}$ decomposition
trivially resembling that of the normal $3/2^-$ state, that is, with its main
contribution coming from the same unique $0\hw$ irrep, while the $1/2^-_2$ state
has a $\grpsu{3}$ decomposition closely resembling that of the intruder $3/2^-$
state, with its main contributions from the two maximally deformed $2\hw$
irreps.  Maximal mixing then occurs between $\Nmax=8$~(center) and $10$.  (A
distance of closest approach of $\approx 2\,\MeV$ here implies a mixing
matrix element half this size~\cite{casten2000:ns}, or $\approx 1\,\MeV$.)  By $\Nmax=12$~(right), an avoided
crossing has transpired, and the \textit{lower} $1/2^-$ state now has the larger
intruder component.

Since the convergence of energies is not complete in even the largest of these
calcluations, we can anticipate that the $3/2^-$ states will continue to mix
more strongly with increasing $\Nmax$, as the intruder state $3/2^-$ continues
to descend in energy towards the normal $3/2^-$ ``ground state''.  (The degree
of mixing seen in Fig.~\ref{fig:li11decomp-wedge-normal} suggests a mixing
matrix element again of $\approx1\,\MeV$ for the $3/2^-$ states as well.)  Similarly,
the mixing of the normal and intruder $1/2^-$ states will continue to decrease,
as the intruder state moves further below the normal state.

The mixing amplitudes arising in an avoided crossing in general depend
sensitively upon both the strength of the mixing matrix element and the
separation in energy (energy denominator) of the states before mixing.
Therefore, caution would need to be exercised in moving from these calculated
results to quantitative predictions for the mixing as physically realized in the
states of $\isotope[11]{Li}$.

However, the qualitative picture is clear, that the NCSM calculations suggest
strong mixing, up to and including the maximal mixing indicated by experiment (for the ground state).
Moreover, despite the large scale of the numerical calculation and \textit{a priori}
complex nature of the microscopic wave functions, the final picture of the
structures entering into this mixing is fairly simple: the normal state follows
the naive spherical shell-model expectations, while the intruder is both mostly
$2\hw$ (ostensibly two-particle, two-hole) and well-deformed.

Given the well-known parity inversion in neighboring $\isotope[11]{Be}$,
discussed earlier, it is not surprising that low-lying states of non-normal
(negative) parity are found in the calculations for $\isotope[11]{Li}$ (green,
open symbols in Fig.~\ref{fig:li11ex}), as well, starting with closely spaced
$5/2^+$ and $3/2^+$ states.  Again, not surprisingly, for the N3LO interaction
[Fig.~\ref{fig:li11ex}(a)], although the convergence with $\Nmax$ is nowhere
near as steep as for the $2\hw$ intruders, the energies of these $1\hw$ states
continue to change steadily with increasing $\Nmax$, while, with the softer
Daejeon16 interaction [Fig.~\ref{fig:li11ex}(b)], the calculated excitation
energies are relatively independent of $\Nmax$, stabilizing with the $5/2^+$
energy just above $2\,\MeV$.  However, it must be kept in mind that these
excitation energies are calculated with reference to a calculated $3/2^-$ ground
state which is still principally normal in nature even at the highest calculated
$\Nmax$.  As the anticipated avoided crossing of the normal and intruder $3/2^-$ states
progresses, and level repulsion becomes more significant, it can be expected
that the $3/2^-$ ground state energy (eigenvalue) may be depressed, thereby raising the excitation
energies (taken relative to this ground state) of the other states.

While spectroscopic quadrupole moments do not provide as much detailed information
about deformation as do group-theoretical decompositions, they are nonetheless
sensitive to deformation and have the advantages of being familiar
observables, straightforward to define, and easy to compute.  Here we take the
dimensionless ratio $Q/r^2$, \textit{i.e.}, the quadrupole moment normalized to
the mean square radius (or, equivalently, to within a prefactor, monopole moment).  This ratio is more rapidly
convergent in NCSM calculations than either the quadrupole moment or mean square
radius individually~\cite{caprio2022:emnorm,caprio2025:emnorm2-part1} and is the
more appropriate measure of deformation for an axially symmetric rotational
nucleus~\cite{bohr1998:v2} (see, \textit{e.g.}, Sec.~II\,C of Ref.~\cite{caprio2025:emnorm2-part1}).  Electromagnetic
measurements of the quadrupole moment (and charge radius) are, of course, are
only directly sensitive to the proton density distribution.  However, in
Fig.~\ref{fig:li11q}, we examine this ratio both for the proton observables (and
thus proton deformation) and for the neutron observables (and thus neutron
deformation), a distinction which turns out to be informative.

Take first the proton ratio $Q_p/r_p^2$ [Fig.~\ref{fig:li11q} (left)].  (Note
that the calculated quadrupole moment is negative, for both the ground and
excited $3/2^-$ states, so increasing magnitude goes downward in the plot.)  For
the ground state, a negative sign for the quadrupole moment is as expected for a
single nucleon outside a closed shell (\textit{e.g.},
Refs.~\cite{suhonen2007:nucleons-nucleus,rowe2010:rowanwood}).  For the excited
$3/2^-$ state, if it is taken to be a member of a $K=1/2$ band (with inverted
spin orderings due to Coriolis staggering), such a negative (spectroscopic)
quadrupole moment might be associated with prolate intrinsic deformation
(\textit{e.g.}, Fig.~1 of Ref.~\cite{maris2015:berotor2}), at least, to the
extent that it might be meaningful to associate a ``deformation'' to the density
distribution generated by just three protons.  Regardless, the proton $Q/r^2$ is
nearly the same for the normal and intruder states ($Q_p/r_p^2\approx -0.6$);
this at least suggests the proton structure is not very different between the
two.  Our results, in the calculations with either the N3LO or Daejeon16
interaction, are in close agreement with
experiment~\cite{stone2016:e2-moments,angeli2013:charge-radii} (solid square in
Fig.~\ref{fig:li11q}).  (Previous NCSM calculations obtained similar values of
$Q_p/r_p^2$~\cite{PhysRevC.57.3119,PhysRevC.79.021303}, including with
the Daejeon16 interaction~\cite{caprio2022:emnorm}.)

The calculated neutron quadrupole moment ratio $Q_n/r_n^2$ [Fig.~\ref{fig:li11q}
  (right)] is much smaller that the corresponding proton ratio for the normal
state ($Q_n/r_n^2\approx-0.2$), as one might expect for small deviations from a
spherical filled neutron shell, but much larger that the corresponding proton
ratio for the intruder state ($Q_n/r_n^2\approx -1.3$), as expected for a more
highly deformed state arising from excitations of neutrons across the shell
closure.  At least, this is the behavior of the observables for the N3LO
calculations [Fig.~\ref{fig:li11q} (b)].

However, in the Daejeon16 calculations [Fig.~\ref{fig:li11q}(d)], the situation
is more subtle, as we might expect in the presence of mixing.  For the ground
state, the neutron quadrupole moment starts small, at low $\Nmax$, as in the
N3LO calculation, but then it steadily climbs, from $\Nmax=4$ onward.
For the excited $3/2^-$ state, the neutron quadrupole moment attains a large
value at low $\Nmax$, as in the N3LO calculation, but then turns over and starts
declining in magnitude after $\Nmax=8$.  This reflects an intruder component (with
large neutron deformation) mixing into the ground state, and a normal component
(with negligible neutron deformation) mixing into the excited state.

\begin{figure}
  \includegraphics[width=\hsize]{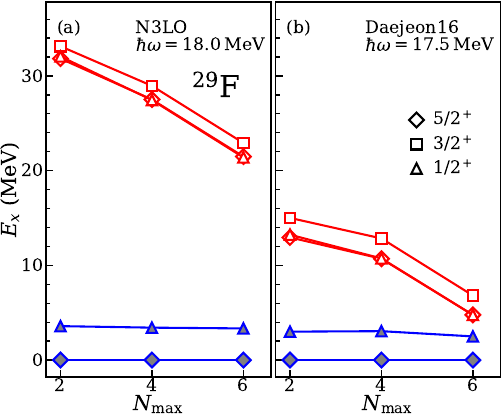}
  \caption{Relative energies for the normal $1/2^+$ and $5/2^+$ (blue, shaded
    symbols) and first intruder $1/2^+$ through $5/2^+$ (red, open symbols)
    levels of $\isotope[29]{F}$, for (a)~a chiral N3LO interaction and (b)~the
    Daejeon16 interaction, shown as functions of $\Nmax$ (at fixed $\hw$, as
    indicated).  }
  \label{fig:f29ex}
\end{figure}
\begin{figure}
  \begin{minipage}{\ifproofpre{1.00}{0.7}\hsize}
    \includegraphics[width=0.49\hsize]{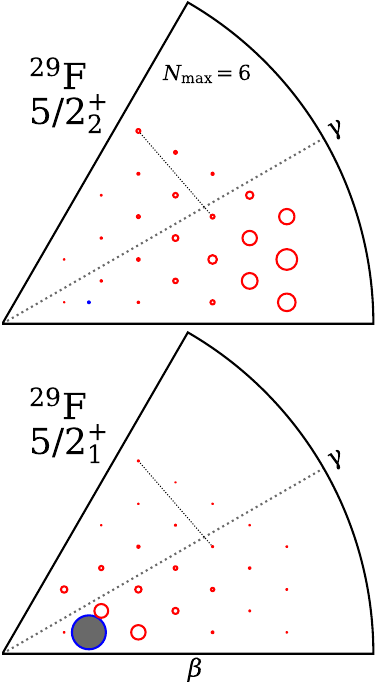}
    \hfill
    \includegraphics[width=0.49\hsize]{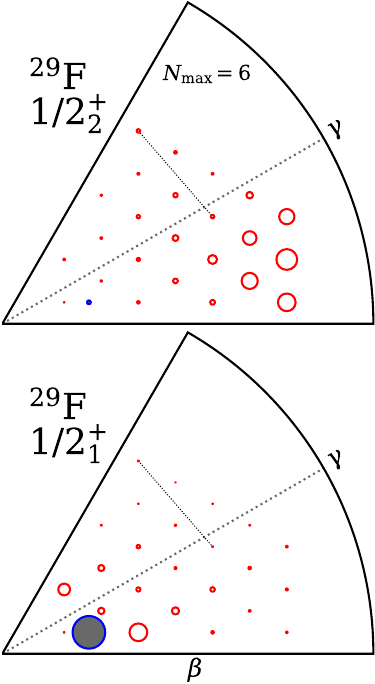}
  \end{minipage}
  \caption{Decompositions by $\grpsu{3}$ jointly with $\Nex$ for the
    first two $5/2^+$~(left) and $1/2^+$~(right) levels of $\isotope[29]{F}$, for
    the Daejeon16 interaction, in calculations with $\Nmax=6$ and
    $\hw=17.5\,\MeV$.  See Fig.~\ref{fig:li11decomp-wedge-normal} caption for
    further explanation of plot contents.}
  \label{fig:f29decomp-wedge-normal}
\end{figure}
\begin{figure}
  \includegraphics[width=1.0\hsize]{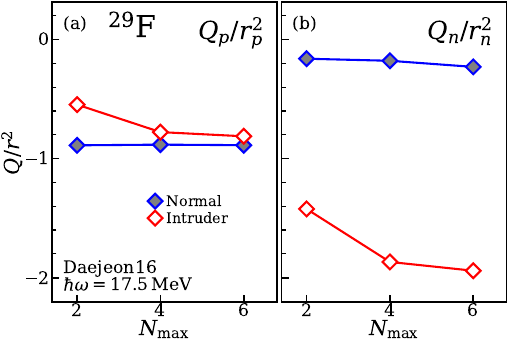}
  \caption{Dimensionless ratio $Q/r^2$ for the normal (blue, shaded symbols) and
    intruder (red, open symbols) $5/2^+$ levels of $\isotope[29]{F}$, for the
    protons~(left) and neutrons~(right), calculated with the Daejeon16
    interaction.  Calculated values are shown as functions of $\Nmax$ (at fixed
    $\hw$, as indicated).  }
  \label{fig:29fq}
\end{figure}

\section{The $N=20$ island of inversion: $\isotope[29]{F}$}
\label{sec:29f}

The nuclide $\isotope[29]{F}$ may, as we have noted, be viewed as a cousin to
$\isotope[11]{Li}$, shifted up by a major oscillator shell. It is another halo
nucleus~\cite{PhysRevLett.124.222504} on an island of
inversion~\cite{fortunato202029f}.  Based at least on model arguments, though
not direct experimental evidence, it is also thought to be
deformed~\cite{macchiavelli2017:29f-nilsson-rotation-aligned,PhysRevC.104.014307}.  The largest NCSM
model space we can presently reach for this nucleus is only $\Nmax=6$, with an
$M$-scheme dimension of $4.7 \times 10^9$.  Nonetheless, the calculations are
informative, since we can see trends which closely parallel those found above
for $\isotope[11]{Li}$.

The calculated energy levels, for the same two interactions (chiral N3LO and
Daejeon16), are shown in Fig.~\ref{fig:f29ex}, again highlighting their evolution
with $\Nmax$, at fixed $\hw$ (here, $\hw=18\,\MeV$ and $17.5\,\MeV$,
respectively, again chosen to approximately minimize the calculated ground state energy for each interaction).  The normal levels (blue, shaded symbols) come lowest, as
expected.  These are now a $5/2^+$ ground state and $1/2^+$ excited state,
corresponding to finding the valence proton in the $0d_{5/2}$ and $1s_{1/2}$
orbitals, respectively (a $3/2^+$ level corresponding to the $0d_{3/2}$
orbital does also arise, but somewhat higher in energy, at $\gtrsim10\,\MeV$, and has been omitted for
clarity).  Then several more closely spaced intruder levels (red, open
symbols) start at much higher energy, but swoop down with increasing $\Nmax$.

Once again, the intruder levels begin not quite so high in energy when
calculated with the softer Daejeon16 interaction [Fig.~\ref{fig:f29ex}(b)]~---
in fact, the starting energy of $\approx13\,\MeV$, for $\Nmax=2$, is far below
the twice $\hw$ one might naively expect from the shell model.  Moreover,
once again, the intruder levels approach the low-lying normal levels much sooner
with the Daejon16 interaction.

The picture of a spherical normal state and a deformed intruder is again
supported by the decompositions into $\grpsu{3}$ irreps, shown in
Fig.~\ref{fig:f29decomp-wedge-normal} for the first two $5/2^+$ states~(left), including the ground state, and the first two $1/2^+$ states~(right). The normal state
in each case (at bottom) is, necessarily, dominated by
the unique $0\hw$ irrep $(2,0)$, corresponding to a single proton in the $sd$
shell, with a closed neutron shell.  The decomposition of the intruder state in each case (at top) is more fragmented than for
$\isotope[11]{Li}$ [Fig.~\ref{fig:li11decomp-wedge-normal}].  The strongest
contributions are distributed over several of these most highly deformed $2\hw$ (red, open
circles, at right in $\beta$-$\gamma$ plot), centered on the $(9,2)$ irrep.
These irreps cover a range of triaxiality, all on the prolate side of
$\gamma=30^\circ$.  [The more triaxial $(8,4)$ leading irrep gives a slightly
  smaller contribution.]  The spread over irreps is perhaps unsurprising,
considering the lore that spin-orbit splitting, which breaks $\grpsu{3}$,
increases with mass number $A$~--- suggesting that a more
appropriate $\grpsu{3}$ description in such nuclei might actually be that of quasi-$\grpsu{3}$~\cite{zuker1995:shell-rotation-quasi-su3,RevModPhys.77.427,nowacki2021:neutron-rich},
rather than Elliott's original harmonic oscillator
$\grpsu{3}$.
Moreover, from comparing Fig.~\ref{fig:f29decomp-wedge-normal} with
Fig.~\ref{fig:li11decomp-wedge-normal}, it is also apparent that there is simply a
greater density of possible $2\hw$ $\grpsu{3}$ irreps for the nucleus to ``choose''
between, within a similar region of the Bohr $\beta$-$\gamma$ deformation space.

With increasing $\Nmax$, the intruder $1/2^+$ is the first to closely approach
its normal counterpart (upward triangles in Fig.~\ref{fig:f29ex}).  Indeed, with
the Daejeon16 interaction, already by $\Nmax=6$, the two approach to within
$\approx2\,\MeV$.  However, such strong mixing as was found for the $1/2^-$
states in $\isotope[11]{Li}$ at similar separation [Fig.~\ref{fig:li11ex}(b)] is
not yet seen here.  Comparing the decompositions of the $1/2^+$ states
[Fig.~\ref{fig:f29decomp-wedge-normal} (right)] indicates a mixing of only $\approx
2\%$, which, for this energy separation, indicates a mixing matrix element
between the normal and intruder states modestly smaller (at $\approx 0.3 \,\MeV$) 
than observed above for $\isotope[11]{Li}$.  Nonetheless, as the intruder
energies continue to fall with $\Nmax$, we can anticipate mixing to increase,
again yielding significant intruder contributions to the ground state or,
depending how far the avoided crossing progresses, an intruder
ground state.

We also compute the dimensionless ratios of the form $Q/r^2$ for proton and
neutron quadrupole moments for $\isotope[29]{F}$, shown in
Fig.~\ref{fig:29fq}. Unsurprisingly, the behavior here is, again, markedly
similar to that seen above for $\isotope[11]{Li}$ in Fig.~\ref{fig:li11q}.

While taking $\isotope[29]{F}$ to higher $\Nmax$ is beyond current capabilities,
we already have strong evidence for the structural parallels to
$\isotope[11]{Li}$.  Once again, for this island-of-inversion nuclide,
deformation-driven intruders are predicted to give rise to shape coexistence in
the low-lying spectrum.

\section{Summary}

In \textit{ab initio} NCSM calculations, while intruder states descend towards
the spherical normal states with increasing model space ($\Nmax$), convergence
is slow, and intruders have thus historically been a challenge to reproduce in
the NCSM.  However, as illustrated in the present calculations, a sufficiently
soft interaction, such as Daejeon16, can provide a marked improvement in
convergence.

We have carried out large no-core shell model calculations of the
island-of-inversion nuclides $\isotope[11]{Li}$ and $\isotope[29]{F}$, using two
interactions, a chiral N3LO interation and the phase-shift-equivalent
interaction Daejeon16.  (More recently reported NCSM calculations for
$\isotope[11]{Li}$~\cite{navratil2026:halo-ab-initio-halo25}, using a different
chiral effective theory interaction including a three-body force, yield
qualitatively consistent convergence patterns and spectra.)  In a naive
spherical shell model picture, these nuclides should have very simple structure:
spherical filled neutron shells and one proton outside filled proton
shells. Instead, the ground state and other low-lying states are either
primarily deformed two-particle, two-hole
intruders or at least undergoing significant mixing with such intruders.  Interestingly, the intruder states are ``simple'' in a
group-theoretical framework, namely, dominated by a few $\grpsu{3}$ irreps,
representing the largest possible deformation within the $2\hw$ space, which, in the present examples,
also leads to triaxial tendencies.

\begin{acknowledgments}
This material is based upon work supported by the U.S.~Department of Energy, Office of Science, Office of Nuclear Physics, under Award Numbers DE-FG02-95ER40934 and DE-FG02-03ER41272.  
This research used resources of the National Energy Research Scientific Computing Center (NERSC), a DOE Office of Science User Facility supported by the Office of Science of the U.S.~Department of Energy under Contract No.~DE-AC02-05CH11231, using NERSC awards NP-ERCAP0023497 and NP-ERCAP0031993. Part of this
research was enabled by computational resources supported by a generous gift to
SDSU from John Oldham.
\end{acknowledgments}


\begin{thebibliography}{85}\makeatletter
\providecommand \@ifxundefined [1]{\@ifx{#1\undefined}
}\providecommand \@ifnum [1]{\ifnum #1\expandafter \@firstoftwo
 \else \expandafter \@secondoftwo
 \fi
}\providecommand \@ifx [1]{\ifx #1\expandafter \@firstoftwo
 \else \expandafter \@secondoftwo
 \fi
}\providecommand \natexlab [1]{#1}\providecommand \enquote  [1]{``#1''}\providecommand \bibnamefont  [1]{#1}\providecommand \bibfnamefont [1]{#1}\providecommand \citenamefont [1]{#1}\providecommand \href@noop [0]{\@secondoftwo}\providecommand \href [0]{\begingroup \@sanitize@url \@href}\providecommand \@href[1]{\@@startlink{#1}\@@href}\providecommand \@@href[1]{\endgroup#1\@@endlink}\providecommand \@sanitize@url [0]{\catcode `\\12\catcode `\$12\catcode
  `\&12\catcode `\#12\catcode `\^12\catcode `\_12\catcode `\%12\relax}\providecommand \@@startlink[1]{}\providecommand \@@endlink[0]{}\providecommand \url  [0]{\begingroup\@sanitize@url \@url }\providecommand \@url [1]{\endgroup\@href {#1}{\urlprefix }}\providecommand \urlprefix  [0]{URL }\providecommand \Eprint [0]{\href }\providecommand \doibase [0]{https://doi.org/}\providecommand \selectlanguage [0]{\@gobble}\providecommand \bibinfo  [0]{\@secondoftwo}\providecommand \bibfield  [0]{\@secondoftwo}\providecommand \translation [1]{[#1]}\providecommand \BibitemOpen [0]{}\providecommand \bibitemStop [0]{}\providecommand \bibitemNoStop [0]{.\EOS\space}\providecommand \EOS [0]{\spacefactor3000\relax}\providecommand \BibitemShut  [1]{\csname bibitem#1\endcsname}\let\auto@bib@innerbib\@empty
\bibitem [{\citenamefont {Nilsson}(1955)}]{nilsson1955:model-NUMBER-AS-PAGE}\BibitemOpen
  \bibfield  {author} {\bibinfo {author} {\bibfnamefont {S.~G.}\ \bibnamefont
  {Nilsson}},\ }\href@noop {} {\bibfield  {journal} {\bibinfo  {journal} {Mat.
  Fys. Medd. Dan. Vid. Selsk.}\ }\textbf {\bibinfo {volume} {29}},\ \bibinfo
  {pages} {16} (\bibinfo {year} {1955})}\BibitemShut {NoStop}\bibitem [{\citenamefont {Lawson}(1980)}]{lawson1980:shell}\BibitemOpen
  \bibfield  {author} {\bibinfo {author} {\bibfnamefont {R.~D.}\ \bibnamefont
  {Lawson}},\ }\href@noop {} {\emph {\bibinfo {title} {Theory of the Nuclear
  Shell Model}}}\ (\bibinfo  {publisher} {Clarendon Press},\ \bibinfo {address}
  {Oxford},\ \bibinfo {year} {1980})\BibitemShut {NoStop}\bibitem [{\citenamefont
  {Elliott}(1958{\natexlab{a}})}]{elliott1958:su3-part1}\BibitemOpen
  \bibfield  {author} {\bibinfo {author} {\bibfnamefont {J.~P.}\ \bibnamefont
  {Elliott}},\ }\href {https://doi.org/10.1098/rspa.1958.0072} {\bibfield
  {journal} {\bibinfo  {journal} {Proc. R. Soc. London A}\ }\textbf {\bibinfo
  {volume} {245}},\ \bibinfo {pages} {128} (\bibinfo {year}
  {1958}{\natexlab{a}})}\BibitemShut {NoStop}\bibitem [{\citenamefont
  {Elliott}(1958{\natexlab{b}})}]{elliott1958:su3-part2}\BibitemOpen
  \bibfield  {author} {\bibinfo {author} {\bibfnamefont {J.~P.}\ \bibnamefont
  {Elliott}},\ }\href {https://doi.org/10.1098/rspa.1958.0101} {\bibfield
  {journal} {\bibinfo  {journal} {Proc. R. Soc. London A}\ }\textbf {\bibinfo
  {volume} {245}},\ \bibinfo {pages} {562} (\bibinfo {year}
  {1958}{\natexlab{b}})}\BibitemShut {NoStop}\bibitem [{\citenamefont {Elliott}\ and\ \citenamefont
  {Harvey}(1963)}]{elliott1963:su3-part3}\BibitemOpen
  \bibfield  {author} {\bibinfo {author} {\bibfnamefont {J.~P.}\ \bibnamefont
  {Elliott}}\ and\ \bibinfo {author} {\bibfnamefont {M.}~\bibnamefont
  {Harvey}},\ }\href {https://doi.org/10.1098/rspa.1963.0071} {\bibfield
  {journal} {\bibinfo  {journal} {Proc. R. Soc. London A}\ }\textbf {\bibinfo
  {volume} {272}},\ \bibinfo {pages} {557} (\bibinfo {year}
  {1963})}\BibitemShut {NoStop}\bibitem [{\citenamefont {Elliott}\ and\ \citenamefont
  {Wilsdon}(1968)}]{elliott1968:su3-part4}\BibitemOpen
  \bibfield  {author} {\bibinfo {author} {\bibfnamefont {J.~P.}\ \bibnamefont
  {Elliott}}\ and\ \bibinfo {author} {\bibfnamefont {C.~E.}\ \bibnamefont
  {Wilsdon}},\ }\href {https://doi.org/10.1098/rspa.1968.0033} {\bibfield
  {journal} {\bibinfo  {journal} {Proc. R. Soc. London A}\ }\textbf {\bibinfo
  {volume} {302}},\ \bibinfo {pages} {509} (\bibinfo {year}
  {1968})}\BibitemShut {NoStop}\bibitem [{\citenamefont {Harvey}(1968)}]{harvey1968:su3-shell}\BibitemOpen
  \bibfield  {author} {\bibinfo {author} {\bibfnamefont {M.}~\bibnamefont
  {Harvey}},\ }\bibinfo {title} {The nuclear $\mathit{SU}_3$ model},\ in\ \href
  {https://doi.org/10.1007/978-1-4757-0103-6_2} {\emph {\bibinfo {booktitle}
  {Advances in Nuclear Physics}}},\ Vol.~\bibinfo {volume} {1},\ \bibinfo
  {editor} {edited by\ \bibinfo {editor} {\bibfnamefont {M.}~\bibnamefont
  {Baranger}}\ and\ \bibinfo {editor} {\bibfnamefont {E.}~\bibnamefont
  {Vogt}}}\ (\bibinfo  {publisher} {Plenum},\ \bibinfo {address} {New York},\
  \bibinfo {year} {1968})\ p.~\bibinfo {pages} {67}\BibitemShut {NoStop}\bibitem [{\citenamefont {Van~Isacker}(2011)}]{vanisacker2011:elliott-legacy}\BibitemOpen
  \bibfield  {author} {\bibinfo {author} {\bibfnamefont {P.}~\bibnamefont
  {Van~Isacker}},\ }\href {https://doi.org/10.1016/j.nuclphysa.2010.12.007}
  {\bibfield  {journal} {\bibinfo  {journal} {Nucl. Phys. A}\ }\textbf
  {\bibinfo {volume} {850}},\ \bibinfo {pages} {157} (\bibinfo {year}
  {2011})}\BibitemShut {NoStop}\bibitem [{\citenamefont {Heyde}\ and\ \citenamefont
  {Wood}(2011)}]{heyde2011:shape-coexistence}\BibitemOpen
  \bibfield  {author} {\bibinfo {author} {\bibfnamefont {K.}~\bibnamefont
  {Heyde}}\ and\ \bibinfo {author} {\bibfnamefont {J.~L.}\ \bibnamefont
  {Wood}},\ }\href {https://doi.org/10.1103/RevModPhys.83.1467} {\bibfield
  {journal} {\bibinfo  {journal} {Rev. Mod. Phys.}\ }\textbf {\bibinfo {volume}
  {83}},\ \bibinfo {pages} {1467} (\bibinfo {year} {2011})}\BibitemShut
  {NoStop}\bibitem [{\citenamefont {Freer}\ and\ \citenamefont
  {Fynbo}(2014)}]{freer2014:12c-hoyle}\BibitemOpen
  \bibfield  {author} {\bibinfo {author} {\bibfnamefont {M.}~\bibnamefont
  {Freer}}\ and\ \bibinfo {author} {\bibfnamefont {H.~O.~U.}\ \bibnamefont
  {Fynbo}},\ }\href {https://doi.org/10.1016/j.ppnp.2014.06.001} {\bibfield
  {journal} {\bibinfo  {journal} {Prog. Part. Nucl. Phys.}\ }\textbf {\bibinfo
  {volume} {78}},\ \bibinfo {pages} {1} (\bibinfo {year} {2014})}\BibitemShut
  {NoStop}\bibitem [{\citenamefont {Epelbaum}\ \emph {et~al.}(2011)\citenamefont
  {Epelbaum}, \citenamefont {Krebs}, \citenamefont {Lee},\ and\ \citenamefont
  {Mei\ss{}ner}}]{PhysRevLett.106.192501}\BibitemOpen
  \bibfield  {author} {\bibinfo {author} {\bibfnamefont {E.}~\bibnamefont
  {Epelbaum}}, \bibinfo {author} {\bibfnamefont {H.}~\bibnamefont {Krebs}},
  \bibinfo {author} {\bibfnamefont {D.}~\bibnamefont {Lee}},\ and\ \bibinfo
  {author} {\bibfnamefont {U.-G.}\ \bibnamefont {Mei\ss{}ner}},\ }\href
  {https://doi.org/10.1103/PhysRevLett.106.192501} {\bibfield  {journal}
  {\bibinfo  {journal} {Phys. Rev. Lett.}\ }\textbf {\bibinfo {volume} {106}},\
  \bibinfo {pages} {192501} (\bibinfo {year} {2011})}\BibitemShut {NoStop}\bibitem [{\citenamefont {Haxton}\ and\ \citenamefont
  {Johnson}(1990)}]{PhysRevLett.65.1325}\BibitemOpen
  \bibfield  {author} {\bibinfo {author} {\bibfnamefont {W.~C.}\ \bibnamefont
  {Haxton}}\ and\ \bibinfo {author} {\bibfnamefont {C.}~\bibnamefont
  {Johnson}},\ }\href {https://doi.org/10.1103/PhysRevLett.65.1325} {\bibfield
  {journal} {\bibinfo  {journal} {Phys. Rev. Lett.}\ }\textbf {\bibinfo
  {volume} {65}},\ \bibinfo {pages} {1325} (\bibinfo {year}
  {1990})}\BibitemShut {NoStop}\bibitem [{\citenamefont {Warburton}\ \emph {et~al.}(1992)\citenamefont
  {Warburton}, \citenamefont {Brown},\ and\ \citenamefont
  {Millener}}]{warburton1992:16o-shell-4ho}\BibitemOpen
  \bibfield  {author} {\bibinfo {author} {\bibfnamefont {E.~K.}\ \bibnamefont
  {Warburton}}, \bibinfo {author} {\bibfnamefont {B.~A.}\ \bibnamefont
  {Brown}},\ and\ \bibinfo {author} {\bibfnamefont {D.~J.}\ \bibnamefont
  {Millener}},\ }\href {https://doi.org/10.1016/0370-2693(92)91472-L}
  {\bibfield  {journal} {\bibinfo  {journal} {Phys. Lett. B}\ }\textbf
  {\bibinfo {volume} {293}},\ \bibinfo {pages} {7} (\bibinfo {year}
  {1992})}\BibitemShut {NoStop}\bibitem [{\citenamefont {Caurier}\ \emph {et~al.}(2007)\citenamefont
  {Caurier}, \citenamefont {Men\'endez}, \citenamefont {Nowacki},\ and\
  \citenamefont {Poves}}]{PhysRevC.75.054317}\BibitemOpen
  \bibfield  {author} {\bibinfo {author} {\bibfnamefont {E.}~\bibnamefont
  {Caurier}}, \bibinfo {author} {\bibfnamefont {J.}~\bibnamefont {Men\'endez}},
  \bibinfo {author} {\bibfnamefont {F.}~\bibnamefont {Nowacki}},\ and\ \bibinfo
  {author} {\bibfnamefont {A.}~\bibnamefont {Poves}},\ }\href
  {https://doi.org/10.1103/PhysRevC.75.054317} {\bibfield  {journal} {\bibinfo
  {journal} {Phys. Rev. C}\ }\textbf {\bibinfo {volume} {75}},\ \bibinfo
  {pages} {054317} (\bibinfo {year} {2007})}\BibitemShut {NoStop}\bibitem [{\citenamefont {Warburton}\ \emph {et~al.}(1990)\citenamefont
  {Warburton}, \citenamefont {Becker},\ and\ \citenamefont
  {Brown}}]{PhysRevC.41.1147}\BibitemOpen
  \bibfield  {author} {\bibinfo {author} {\bibfnamefont {E.~K.}\ \bibnamefont
  {Warburton}}, \bibinfo {author} {\bibfnamefont {J.~A.}\ \bibnamefont
  {Becker}},\ and\ \bibinfo {author} {\bibfnamefont {B.~A.}\ \bibnamefont
  {Brown}},\ }\href {https://doi.org/10.1103/PhysRevC.41.1147} {\bibfield
  {journal} {\bibinfo  {journal} {Phys. Rev. C}\ }\textbf {\bibinfo {volume}
  {41}},\ \bibinfo {pages} {1147} (\bibinfo {year} {1990})}\BibitemShut
  {NoStop}\bibitem [{\citenamefont {Caurier}\ \emph {et~al.}(2014)\citenamefont
  {Caurier}, \citenamefont {Nowacki},\ and\ \citenamefont
  {Poves}}]{PhysRevC.90.014302}\BibitemOpen
  \bibfield  {author} {\bibinfo {author} {\bibfnamefont {E.}~\bibnamefont
  {Caurier}}, \bibinfo {author} {\bibfnamefont {F.}~\bibnamefont {Nowacki}},\
  and\ \bibinfo {author} {\bibfnamefont {A.}~\bibnamefont {Poves}},\ }\href
  {https://doi.org/10.1103/PhysRevC.90.014302} {\bibfield  {journal} {\bibinfo
  {journal} {Phys. Rev. C}\ }\textbf {\bibinfo {volume} {90}},\ \bibinfo
  {pages} {014302} (\bibinfo {year} {2014})}\BibitemShut {NoStop}\bibitem [{\citenamefont {Talmi}\ and\ \citenamefont
  {Unna}(1960)}]{PhysRevLett.4.469}\BibitemOpen
  \bibfield  {author} {\bibinfo {author} {\bibfnamefont {I.}~\bibnamefont
  {Talmi}}\ and\ \bibinfo {author} {\bibfnamefont {I.}~\bibnamefont {Unna}},\
  }\href {https://doi.org/10.1103/PhysRevLett.4.469} {\bibfield  {journal}
  {\bibinfo  {journal} {Phys. Rev. Lett.}\ }\textbf {\bibinfo {volume} {4}},\
  \bibinfo {pages} {469} (\bibinfo {year} {1960})}\BibitemShut {NoStop}\bibitem [{\citenamefont {Bohr}\ and\ \citenamefont
  {Mottelson}(1998)}]{bohr1998:v2}\BibitemOpen
  \bibfield  {author} {\bibinfo {author} {\bibfnamefont {A.}~\bibnamefont
  {Bohr}}\ and\ \bibinfo {author} {\bibfnamefont {B.~R.}\ \bibnamefont
  {Mottelson}},\ }\href {https://doi.org/10.1142/3530} {\emph {\bibinfo {title}
  {Nuclear Structure}}},\ Vol.~\bibinfo {volume} {2}\ (\bibinfo  {publisher}
  {World Scientific},\ \bibinfo {address} {Singapore},\ \bibinfo {year}
  {1998})\BibitemShut {NoStop}\bibitem [{\citenamefont {Hamamoto}\ and\ \citenamefont
  {Shimoura}(2007)}]{hamamoto2007:11be-12be-nilsson}\BibitemOpen
  \bibfield  {author} {\bibinfo {author} {\bibfnamefont {I.}~\bibnamefont
  {Hamamoto}}\ and\ \bibinfo {author} {\bibfnamefont {S.}~\bibnamefont
  {Shimoura}},\ }\href {https://doi.org/10.1088/0954-3899/34/12/015} {\bibfield
   {journal} {\bibinfo  {journal} {J. Phys. G}\ }\textbf {\bibinfo {volume}
  {34}},\ \bibinfo {pages} {2715} (\bibinfo {year} {2007})}\BibitemShut
  {NoStop}\bibitem [{\citenamefont {Macchiavelli}\ \emph {et~al.}(2018)\citenamefont
  {Macchiavelli}, \citenamefont {Crawford}, \citenamefont {Campbell},
  \citenamefont {Clark}, \citenamefont {Cromaz}, \citenamefont {Fallon},
  \citenamefont {Jones}, \citenamefont {Lee},\ and\ \citenamefont
  {Salathe}}]{macchiavelli2018:11be-12be-nilsson}\BibitemOpen
  \bibfield  {author} {\bibinfo {author} {\bibfnamefont {A.~O.}\ \bibnamefont
  {Macchiavelli}}, \bibinfo {author} {\bibfnamefont {H.~L.}\ \bibnamefont
  {Crawford}}, \bibinfo {author} {\bibfnamefont {C.~M.}\ \bibnamefont
  {Campbell}}, \bibinfo {author} {\bibfnamefont {R.~M.}\ \bibnamefont {Clark}},
  \bibinfo {author} {\bibfnamefont {M.}~\bibnamefont {Cromaz}}, \bibinfo
  {author} {\bibfnamefont {P.}~\bibnamefont {Fallon}}, \bibinfo {author}
  {\bibfnamefont {M.~D.}\ \bibnamefont {Jones}}, \bibinfo {author}
  {\bibfnamefont {I.~Y.}\ \bibnamefont {Lee}},\ and\ \bibinfo {author}
  {\bibfnamefont {M.}~\bibnamefont {Salathe}},\ }\href
  {https://doi.org/10.1103/PhysRevC.97.011302} {\bibfield  {journal} {\bibinfo
  {journal} {Phys. Rev. C}\ }\textbf {\bibinfo {volume} {97}},\ \bibinfo
  {pages} {011302} (\bibinfo {year} {2018})}\BibitemShut {NoStop}\bibitem [{\citenamefont {Caurier}\ \emph {et~al.}(2005)\citenamefont
  {Caurier}, \citenamefont {Mart\'{\i}nez-Pinedo}, \citenamefont {Nowacki},
  \citenamefont {Poves},\ and\ \citenamefont {Zuker}}]{RevModPhys.77.427}\BibitemOpen
  \bibfield  {author} {\bibinfo {author} {\bibfnamefont {E.}~\bibnamefont
  {Caurier}}, \bibinfo {author} {\bibfnamefont {G.}~\bibnamefont
  {Mart\'{\i}nez-Pinedo}}, \bibinfo {author} {\bibfnamefont {F.}~\bibnamefont
  {Nowacki}}, \bibinfo {author} {\bibfnamefont {A.}~\bibnamefont {Poves}},\
  and\ \bibinfo {author} {\bibfnamefont {A.~P.}\ \bibnamefont {Zuker}},\ }\href
  {https://doi.org/10.1103/RevModPhys.77.427} {\bibfield  {journal} {\bibinfo
  {journal} {Rev. Mod. Phys.}\ }\textbf {\bibinfo {volume} {77}},\ \bibinfo
  {pages} {427} (\bibinfo {year} {2005})}\BibitemShut {NoStop}\bibitem [{\citenamefont {Otsuka}\ \emph {et~al.}(2020)\citenamefont {Otsuka},
  \citenamefont {Gade}, \citenamefont {Sorlin}, \citenamefont {Suzuki},\ and\
  \citenamefont {Utsuno}}]{otsuka2020:shell-structure}\BibitemOpen
  \bibfield  {author} {\bibinfo {author} {\bibfnamefont {T.}~\bibnamefont
  {Otsuka}}, \bibinfo {author} {\bibfnamefont {A.}~\bibnamefont {Gade}},
  \bibinfo {author} {\bibfnamefont {O.}~\bibnamefont {Sorlin}}, \bibinfo
  {author} {\bibfnamefont {T.}~\bibnamefont {Suzuki}},\ and\ \bibinfo {author}
  {\bibfnamefont {Y.}~\bibnamefont {Utsuno}},\ }\href
  {https://doi.org/10.1103/revmodphys.92.015002} {\bibfield  {journal}
  {\bibinfo  {journal} {Rev. Mod. Phys.}\ }\textbf {\bibinfo {volume} {92}},\
  \bibinfo {pages} {015002} (\bibinfo {year} {2020})}\BibitemShut {NoStop}\bibitem [{\citenamefont {Nowacki}\ \emph {et~al.}(2021)\citenamefont
  {Nowacki}, \citenamefont {Obertelli},\ and\ \citenamefont
  {Poves}}]{nowacki2021:neutron-rich}\BibitemOpen
  \bibfield  {author} {\bibinfo {author} {\bibfnamefont {F.}~\bibnamefont
  {Nowacki}}, \bibinfo {author} {\bibfnamefont {A.}~\bibnamefont {Obertelli}},\
  and\ \bibinfo {author} {\bibfnamefont {A.}~\bibnamefont {Poves}},\ }\href
  {https://doi.org/10.1016/j.ppnp.2021.103866} {\bibfield  {journal} {\bibinfo
  {journal} {Prog. Part. Nucl. Phys.}\ }\textbf {\bibinfo {volume} {120}},\
  \bibinfo {pages} {103866} (\bibinfo {year} {2021})}\BibitemShut {NoStop}\bibitem [{\citenamefont {Navr\'{a}til}\ \emph {et~al.}(2000)\citenamefont
  {Navr\'{a}til}, \citenamefont {Vary},\ and\ \citenamefont
  {Barrett}}]{navratil2000:12c-ncsm}\BibitemOpen
  \bibfield  {author} {\bibinfo {author} {\bibfnamefont {P.}~\bibnamefont
  {Navr\'{a}til}}, \bibinfo {author} {\bibfnamefont {J.~P.}\ \bibnamefont
  {Vary}},\ and\ \bibinfo {author} {\bibfnamefont {B.~R.}\ \bibnamefont
  {Barrett}},\ }\href {https://doi.org/10.1103/PhysRevC.62.054311} {\bibfield
  {journal} {\bibinfo  {journal} {Phys. Rev. C}\ }\textbf {\bibinfo {volume}
  {62}},\ \bibinfo {pages} {054311} (\bibinfo {year} {2000})}\BibitemShut
  {NoStop}\bibitem [{\citenamefont {McCoy}\ \emph {et~al.}(2024)\citenamefont {McCoy},
  \citenamefont {Caprio}, \citenamefont {Maris},\ and\ \citenamefont
  {Fasano}}]{mccoy2024:12be-shape}\BibitemOpen
  \bibfield  {author} {\bibinfo {author} {\bibfnamefont {A.~E.}\ \bibnamefont
  {McCoy}}, \bibinfo {author} {\bibfnamefont {M.~A.}\ \bibnamefont {Caprio}},
  \bibinfo {author} {\bibfnamefont {P.}~\bibnamefont {Maris}},\ and\ \bibinfo
  {author} {\bibfnamefont {P.~J.}\ \bibnamefont {Fasano}},\ }\href
  {https://doi.org/10.1016/j.physletb.2024.138870} {\bibfield  {journal}
  {\bibinfo  {journal} {Phys. Lett. B}\ }\textbf {\bibinfo {volume} {856}},\
  \bibinfo {pages} {138870} (\bibinfo {year} {2024})}\BibitemShut {NoStop}\bibitem [{\citenamefont {Casta{\~n}os}\ \emph {et~al.}(1988)\citenamefont
  {Casta{\~n}os}, \citenamefont {Draayer},\ and\ \citenamefont
  {Leschber}}]{castanos1988:su3-shape}\BibitemOpen
  \bibfield  {author} {\bibinfo {author} {\bibfnamefont {O.}~\bibnamefont
  {Casta{\~n}os}}, \bibinfo {author} {\bibfnamefont {J.~P.}\ \bibnamefont
  {Draayer}},\ and\ \bibinfo {author} {\bibfnamefont {Y.}~\bibnamefont
  {Leschber}},\ }\href {https://doi.org/10.1007/BF01294813} {\bibfield
  {journal} {\bibinfo  {journal} {Z. Phys. A}\ }\textbf {\bibinfo {volume}
  {329}},\ \bibinfo {pages} {33} (\bibinfo {year} {1988})}\BibitemShut
  {NoStop}\bibitem [{\citenamefont {Rowe}\ \emph {et~al.}(2006)\citenamefont {Rowe},
  \citenamefont {Thiamova},\ and\ \citenamefont
  {Wood}}]{rowe2006:coexistence-shell-u3}\BibitemOpen
  \bibfield  {author} {\bibinfo {author} {\bibfnamefont {D.~J.}\ \bibnamefont
  {Rowe}}, \bibinfo {author} {\bibfnamefont {G.}~\bibnamefont {Thiamova}},\
  and\ \bibinfo {author} {\bibfnamefont {J.~L.}\ \bibnamefont {Wood}},\ }\href
  {https://doi.org/10.1103/PhysRevLett.97.202501} {\bibfield  {journal}
  {\bibinfo  {journal} {Phys. Rev. Lett.}\ }\textbf {\bibinfo {volume} {97}},\
  \bibinfo {pages} {202501} (\bibinfo {year} {2006})}\BibitemShut {NoStop}\bibitem [{\citenamefont {Dreyfuss}\ \emph {et~al.}(2013)\citenamefont
  {Dreyfuss}, \citenamefont {Launey}, \citenamefont {Dytrych}, \citenamefont
  {Draayer},\ and\ \citenamefont {Bahri}}]{dreyfuss2013:12c-sp-rotation}\BibitemOpen
  \bibfield  {author} {\bibinfo {author} {\bibfnamefont {A.~C.}\ \bibnamefont
  {Dreyfuss}}, \bibinfo {author} {\bibfnamefont {K.~D.}\ \bibnamefont
  {Launey}}, \bibinfo {author} {\bibfnamefont {T.}~\bibnamefont {Dytrych}},
  \bibinfo {author} {\bibfnamefont {J.~P.}\ \bibnamefont {Draayer}},\ and\
  \bibinfo {author} {\bibfnamefont {C.}~\bibnamefont {Bahri}},\ }\href
  {https://doi.org/10.1016/j.physletb.2013.10.048} {\bibfield  {journal}
  {\bibinfo  {journal} {Phys. Lett. B}\ }\textbf {\bibinfo {volume} {727}},\
  \bibinfo {pages} {511} (\bibinfo {year} {2013})}\BibitemShut {NoStop}\bibitem [{\citenamefont {Rowe}(2020)}]{rowe2020:shape-coexistence-algebraic}\BibitemOpen
  \bibfield  {author} {\bibinfo {author} {\bibfnamefont {D.~J.}\ \bibnamefont
  {Rowe}},\ }\href {https://doi.org/10.1103/PhysRevC.101.054301} {\bibfield
  {journal} {\bibinfo  {journal} {Phys. Rev. C}\ }\textbf {\bibinfo {volume}
  {101}},\ \bibinfo {pages} {054301} (\bibinfo {year} {2020})}\BibitemShut
  {NoStop}\bibitem [{\citenamefont {Caurier}\ \emph {et~al.}(2001)\citenamefont
  {Caurier}, \citenamefont {Navr\'atil}, \citenamefont {Ormand},\ and\
  \citenamefont {Vary}}]{PhysRevC.64.051301}\BibitemOpen
  \bibfield  {author} {\bibinfo {author} {\bibfnamefont {E.}~\bibnamefont
  {Caurier}}, \bibinfo {author} {\bibfnamefont {P.}~\bibnamefont {Navr\'atil}},
  \bibinfo {author} {\bibfnamefont {W.~E.}\ \bibnamefont {Ormand}},\ and\
  \bibinfo {author} {\bibfnamefont {J.~P.}\ \bibnamefont {Vary}},\ }\href
  {https://doi.org/10.1103/PhysRevC.64.051301} {\bibfield  {journal} {\bibinfo
  {journal} {Phys. Rev. C}\ }\textbf {\bibinfo {volume} {64}},\ \bibinfo
  {pages} {051301} (\bibinfo {year} {2001})}\BibitemShut {NoStop}\bibitem [{\citenamefont {Caurier}\ \emph {et~al.}(2002)\citenamefont
  {Caurier}, \citenamefont {Navr\'atil}, \citenamefont {Ormand},\ and\
  \citenamefont {Vary}}]{PhysRevC.66.024314}\BibitemOpen
  \bibfield  {author} {\bibinfo {author} {\bibfnamefont {E.}~\bibnamefont
  {Caurier}}, \bibinfo {author} {\bibfnamefont {P.}~\bibnamefont {Navr\'atil}},
  \bibinfo {author} {\bibfnamefont {W.~E.}\ \bibnamefont {Ormand}},\ and\
  \bibinfo {author} {\bibfnamefont {J.~P.}\ \bibnamefont {Vary}},\ }\href
  {https://doi.org/10.1103/PhysRevC.66.024314} {\bibfield  {journal} {\bibinfo
  {journal} {Phys. Rev. C}\ }\textbf {\bibinfo {volume} {66}},\ \bibinfo
  {pages} {024314} (\bibinfo {year} {2002})}\BibitemShut {NoStop}\bibitem [{\citenamefont {Navr\'{a}til}\ and\ \citenamefont
  {Ormand}(2003)}]{navratil2003:ncsm-3n}\BibitemOpen
  \bibfield  {author} {\bibinfo {author} {\bibfnamefont {P.}~\bibnamefont
  {Navr\'{a}til}}\ and\ \bibinfo {author} {\bibfnamefont {W.~E.}\ \bibnamefont
  {Ormand}},\ }\href {https://doi.org/10.1103/PhysRevC.68.034305} {\bibfield
  {journal} {\bibinfo  {journal} {Phys. Rev. C}\ }\textbf {\bibinfo {volume}
  {68}},\ \bibinfo {pages} {034305} (\bibinfo {year} {2003})}\BibitemShut
  {NoStop}\bibitem [{\citenamefont {Forss{\'e}n}\ \emph {et~al.}(2005)\citenamefont
  {Forss{\'e}n}, \citenamefont {Navr{\'a}til}, \citenamefont {Ormand},\ and\
  \citenamefont {Caurier}}]{forssen2005:ncsm-9be-11be}\BibitemOpen
  \bibfield  {author} {\bibinfo {author} {\bibfnamefont {C.}~\bibnamefont
  {Forss{\'e}n}}, \bibinfo {author} {\bibfnamefont {P.}~\bibnamefont
  {Navr{\'a}til}}, \bibinfo {author} {\bibfnamefont {W.~E.}\ \bibnamefont
  {Ormand}},\ and\ \bibinfo {author} {\bibfnamefont {E.}~\bibnamefont
  {Caurier}},\ }\href {https://doi.org/10.1103/PhysRevC.71.044312} {\bibfield
  {journal} {\bibinfo  {journal} {Phys. Rev. C}\ }\textbf {\bibinfo {volume}
  {71}},\ \bibinfo {pages} {044312} (\bibinfo {year} {2005})}\BibitemShut
  {NoStop}\bibitem [{\citenamefont {Bogner}\ \emph {et~al.}(2007)\citenamefont {Bogner},
  \citenamefont {Furnstahl},\ and\ \citenamefont {Perry}}]{PhysRevC.75.061001}\BibitemOpen
  \bibfield  {author} {\bibinfo {author} {\bibfnamefont {S.~K.}\ \bibnamefont
  {Bogner}}, \bibinfo {author} {\bibfnamefont {R.~J.}\ \bibnamefont
  {Furnstahl}},\ and\ \bibinfo {author} {\bibfnamefont {R.~J.}\ \bibnamefont
  {Perry}},\ }\href {https://doi.org/10.1103/PhysRevC.75.061001} {\bibfield
  {journal} {\bibinfo  {journal} {Phys. Rev. C}\ }\textbf {\bibinfo {volume}
  {75}},\ \bibinfo {pages} {061001} (\bibinfo {year} {2007})}\BibitemShut
  {NoStop}\bibitem [{\citenamefont {Jurgenson}\ \emph {et~al.}(2011)\citenamefont
  {Jurgenson}, \citenamefont {Navr\'atil},\ and\ \citenamefont
  {Furnstahl}}]{PhysRevC.83.034301}\BibitemOpen
  \bibfield  {author} {\bibinfo {author} {\bibfnamefont {E.~D.}\ \bibnamefont
  {Jurgenson}}, \bibinfo {author} {\bibfnamefont {P.}~\bibnamefont
  {Navr\'atil}},\ and\ \bibinfo {author} {\bibfnamefont {R.~J.}\ \bibnamefont
  {Furnstahl}},\ }\href {https://doi.org/10.1103/PhysRevC.83.034301} {\bibfield
   {journal} {\bibinfo  {journal} {Phys. Rev. C}\ }\textbf {\bibinfo {volume}
  {83}},\ \bibinfo {pages} {034301} (\bibinfo {year} {2011})}\BibitemShut
  {NoStop}\bibitem [{\citenamefont {Caprio}\ \emph {et~al.}(2019)\citenamefont {Caprio},
  \citenamefont {Fasano}, \citenamefont {McCoy}, \citenamefont {Maris},\ and\
  \citenamefont {Vary}}]{caprio2019:bebands-sdanca19}\BibitemOpen
  \bibfield  {author} {\bibinfo {author} {\bibfnamefont {M.~A.}\ \bibnamefont
  {Caprio}}, \bibinfo {author} {\bibfnamefont {P.~J.}\ \bibnamefont {Fasano}},
  \bibinfo {author} {\bibfnamefont {A.~E.}\ \bibnamefont {McCoy}}, \bibinfo
  {author} {\bibfnamefont {P.}~\bibnamefont {Maris}},\ and\ \bibinfo {author}
  {\bibfnamefont {J.~P.}\ \bibnamefont {Vary}},\ }\href@noop {} {\bibfield
  {journal} {\bibinfo  {journal} {Bulg. J. Phys.}\ }\textbf {\bibinfo {volume}
  {46}},\ \bibinfo {pages} {445} (\bibinfo {year} {2019})}\BibitemShut
  {NoStop}\bibitem [{\citenamefont {Caprio}\ \emph {et~al.}(2020)\citenamefont {Caprio},
  \citenamefont {Fasano}, \citenamefont {Maris}, \citenamefont {McCoy},\ and\
  \citenamefont {Vary}}]{caprio2020:bebands}\BibitemOpen
  \bibfield  {author} {\bibinfo {author} {\bibfnamefont {M.~A.}\ \bibnamefont
  {Caprio}}, \bibinfo {author} {\bibfnamefont {P.~J.}\ \bibnamefont {Fasano}},
  \bibinfo {author} {\bibfnamefont {P.}~\bibnamefont {Maris}}, \bibinfo
  {author} {\bibfnamefont {A.~E.}\ \bibnamefont {McCoy}},\ and\ \bibinfo
  {author} {\bibfnamefont {J.~P.}\ \bibnamefont {Vary}},\ }\href
  {https://doi.org/10.1140/epja/s10050-020-00112-0} {\bibfield  {journal}
  {\bibinfo  {journal} {Eur. Phys. J. A}\ }\textbf {\bibinfo {volume} {56}},\
  \bibinfo {pages} {120} (\bibinfo {year} {2020})}\BibitemShut {NoStop}\bibitem [{\citenamefont {Rowe}(1996)}]{rowe1996:sp3r-dynamical-symmetry}\BibitemOpen
  \bibfield  {author} {\bibinfo {author} {\bibfnamefont {D.~J.}\ \bibnamefont
  {Rowe}},\ }\href {https://doi.org/10.1016/0146-6410(96)00058-0} {\bibfield
  {journal} {\bibinfo  {journal} {Prog. Part. Nucl. Phys.}\ }\textbf {\bibinfo
  {volume} {37}},\ \bibinfo {pages} {265} (\bibinfo {year} {1996})}\BibitemShut
  {NoStop}\bibitem [{\citenamefont {McCoy}(2018)}]{mccoy2018:diss}\BibitemOpen
  \bibfield  {author} {\bibinfo {author} {\bibfnamefont {A.~E.}\ \bibnamefont
  {McCoy}},\ }\emph {\bibinfo {title} {\textit{Ab initio} multi-irrep
  symplectic no-core configuration interaction calculations}},\ \href
  {https://doi.org/10.7274/pz50gt57p16} {Ph.D. thesis},\ \bibinfo  {school}
  {University of Notre Dame} (\bibinfo {year} {2018})\BibitemShut {NoStop}\bibitem [{\citenamefont {McCoy}\ \emph {et~al.}(2020)\citenamefont {McCoy},
  \citenamefont {Caprio}, \citenamefont {Dytrych},\ and\ \citenamefont
  {Fasano}}]{mccoy2020:spfamilies}\BibitemOpen
  \bibfield  {author} {\bibinfo {author} {\bibfnamefont {A.~E.}\ \bibnamefont
  {McCoy}}, \bibinfo {author} {\bibfnamefont {M.~A.}\ \bibnamefont {Caprio}},
  \bibinfo {author} {\bibfnamefont {T.}~\bibnamefont {Dytrych}},\ and\ \bibinfo
  {author} {\bibfnamefont {P.~J.}\ \bibnamefont {Fasano}},\ }\href
  {https://doi.org/10.1103/PhysRevLett.125.102505} {\bibfield  {journal}
  {\bibinfo  {journal} {Phys. Rev. Lett.}\ }\textbf {\bibinfo {volume} {125}},\
  \bibinfo {pages} {102505} (\bibinfo {year} {2020})}\BibitemShut {NoStop}\bibitem [{\citenamefont {Dytrych}\ \emph {et~al.}(2020)\citenamefont
  {Dytrych}, \citenamefont {Launey}, \citenamefont {Draayer}, \citenamefont
  {Rowe}, \citenamefont {Wood}, \citenamefont {Rosensteel}, \citenamefont
  {Bahri}, \citenamefont {Langr},\ and\ \citenamefont
  {Baker}}]{dytrych2020:emergent-symmetry}\BibitemOpen
  \bibfield  {author} {\bibinfo {author} {\bibfnamefont {T.}~\bibnamefont
  {Dytrych}}, \bibinfo {author} {\bibfnamefont {K.~D.}\ \bibnamefont {Launey}},
  \bibinfo {author} {\bibfnamefont {J.~P.}\ \bibnamefont {Draayer}}, \bibinfo
  {author} {\bibfnamefont {D.~J.}\ \bibnamefont {Rowe}}, \bibinfo {author}
  {\bibfnamefont {J.~L.}\ \bibnamefont {Wood}}, \bibinfo {author}
  {\bibfnamefont {G.}~\bibnamefont {Rosensteel}}, \bibinfo {author}
  {\bibfnamefont {C.}~\bibnamefont {Bahri}}, \bibinfo {author} {\bibfnamefont
  {D.}~\bibnamefont {Langr}},\ and\ \bibinfo {author} {\bibfnamefont {R.~B.}\
  \bibnamefont {Baker}},\ }\href
  {https://doi.org/10.1103/PhysRevLett.124.042501} {\bibfield  {journal}
  {\bibinfo  {journal} {Phys. Rev. Lett.}\ }\textbf {\bibinfo {volume} {124}},\
  \bibinfo {pages} {042501} (\bibinfo {year} {2020})}\BibitemShut {NoStop}\bibitem [{\citenamefont {Zbikowski}\ \emph {et~al.}(2021)\citenamefont
  {Zbikowski}, \citenamefont {Johnson}, \citenamefont {McCoy}, \citenamefont
  {Caprio},\ and\ \citenamefont {Fasano}}]{zbikowski2021:beyond-elliott}\BibitemOpen
  \bibfield  {author} {\bibinfo {author} {\bibfnamefont {R.}~\bibnamefont
  {Zbikowski}}, \bibinfo {author} {\bibfnamefont {C.~W.}\ \bibnamefont
  {Johnson}}, \bibinfo {author} {\bibfnamefont {A.~E.}\ \bibnamefont {McCoy}},
  \bibinfo {author} {\bibfnamefont {M.~A.}\ \bibnamefont {Caprio}},\ and\
  \bibinfo {author} {\bibfnamefont {P.~J.}\ \bibnamefont {Fasano}},\ }\href
  {https://doi.org/10.1088/1361-6471/abdd8e} {\bibfield  {journal} {\bibinfo
  {journal} {J. Phys. G}\ }\textbf {\bibinfo {volume} {48}},\ \bibinfo {pages}
  {075102} (\bibinfo {year} {2021})}\BibitemShut {NoStop}\bibitem [{\citenamefont {Caprio}\ \emph
  {et~al.}(2022{\natexlab{a}})\citenamefont {Caprio}, \citenamefont {McCoy},
  \citenamefont {Fasano},\ and\ \citenamefont
  {Dytrych}}]{caprio2022:10be-shape-sdanca21}\BibitemOpen
  \bibfield  {author} {\bibinfo {author} {\bibfnamefont {M.~A.}\ \bibnamefont
  {Caprio}}, \bibinfo {author} {\bibfnamefont {A.~E.}\ \bibnamefont {McCoy}},
  \bibinfo {author} {\bibfnamefont {P.~J.}\ \bibnamefont {Fasano}},\ and\
  \bibinfo {author} {\bibfnamefont {T.}~\bibnamefont {Dytrych}},\ }\href
  {https://doi.org/10.55318/bgjp.2022.49.1.057} {\bibfield  {journal} {\bibinfo
   {journal} {Bulg. J. Phys.}\ }\textbf {\bibinfo {volume} {49}},\ \bibinfo
  {pages} {57} (\bibinfo {year} {2022}{\natexlab{a}})}\BibitemShut {NoStop}\bibitem [{\citenamefont {Sargsyan}\ \emph {et~al.}(2022)\citenamefont
  {Sargsyan}, \citenamefont {Launey}, \citenamefont {Burkey}, \citenamefont
  {Gallant}, \citenamefont {Scielzo}, \citenamefont {Savard}, \citenamefont
  {Mercenne}, \citenamefont {Dytrych}, \citenamefont {Langr}, \citenamefont
  {Varriano}, \citenamefont {Longfellow}, \citenamefont {Hirsh},\ and\
  \citenamefont {Draayer}}]{PhysRevLett.128.202503}\BibitemOpen
  \bibfield  {author} {\bibinfo {author} {\bibfnamefont {G.~H.}\ \bibnamefont
  {Sargsyan}}, \bibinfo {author} {\bibfnamefont {K.~D.}\ \bibnamefont
  {Launey}}, \bibinfo {author} {\bibfnamefont {M.~T.}\ \bibnamefont {Burkey}},
  \bibinfo {author} {\bibfnamefont {A.~T.}\ \bibnamefont {Gallant}}, \bibinfo
  {author} {\bibfnamefont {N.~D.}\ \bibnamefont {Scielzo}}, \bibinfo {author}
  {\bibfnamefont {G.}~\bibnamefont {Savard}}, \bibinfo {author} {\bibfnamefont
  {A.}~\bibnamefont {Mercenne}}, \bibinfo {author} {\bibfnamefont
  {T.}~\bibnamefont {Dytrych}}, \bibinfo {author} {\bibfnamefont
  {D.}~\bibnamefont {Langr}}, \bibinfo {author} {\bibfnamefont
  {L.}~\bibnamefont {Varriano}}, \bibinfo {author} {\bibfnamefont
  {B.}~\bibnamefont {Longfellow}}, \bibinfo {author} {\bibfnamefont {T.~Y.}\
  \bibnamefont {Hirsh}},\ and\ \bibinfo {author} {\bibfnamefont {J.~P.}\
  \bibnamefont {Draayer}},\ }\href
  {https://doi.org/10.1103/PhysRevLett.128.202503} {\bibfield  {journal}
  {\bibinfo  {journal} {Phys. Rev. Lett.}\ }\textbf {\bibinfo {volume} {128}},\
  \bibinfo {pages} {202503} (\bibinfo {year} {2022})}\BibitemShut {NoStop}\bibitem [{\citenamefont {Tanihata}\ \emph {et~al.}(1985)\citenamefont
  {Tanihata}, \citenamefont {Hamagaki}, \citenamefont {Hashimoto},
  \citenamefont {Shida}, \citenamefont {Yoshikawa}, \citenamefont {Sugimoto},
  \citenamefont {Yamakawa}, \citenamefont {Kobayashi},\ and\ \citenamefont
  {Takahashi}}]{tanihata1985:radii-11li-halo}\BibitemOpen
  \bibfield  {author} {\bibinfo {author} {\bibfnamefont {I.}~\bibnamefont
  {Tanihata}}, \bibinfo {author} {\bibfnamefont {H.}~\bibnamefont {Hamagaki}},
  \bibinfo {author} {\bibfnamefont {O.}~\bibnamefont {Hashimoto}}, \bibinfo
  {author} {\bibfnamefont {Y.}~\bibnamefont {Shida}}, \bibinfo {author}
  {\bibfnamefont {N.}~\bibnamefont {Yoshikawa}}, \bibinfo {author}
  {\bibfnamefont {K.}~\bibnamefont {Sugimoto}}, \bibinfo {author}
  {\bibfnamefont {O.}~\bibnamefont {Yamakawa}}, \bibinfo {author}
  {\bibfnamefont {T.}~\bibnamefont {Kobayashi}},\ and\ \bibinfo {author}
  {\bibfnamefont {N.}~\bibnamefont {Takahashi}},\ }\href
  {https://doi.org/10.1103/PhysRevLett.55.2676} {\bibfield  {journal} {\bibinfo
   {journal} {Phys. Rev. Lett.}\ }\textbf {\bibinfo {volume} {55}},\ \bibinfo
  {pages} {2676} (\bibinfo {year} {1985})}\BibitemShut {NoStop}\bibitem [{\citenamefont {Tanihata}\ \emph {et~al.}(1992)\citenamefont
  {Tanihata}, \citenamefont {Kobayashi}, \citenamefont {Suzuki}, \citenamefont
  {Yoshida}, \citenamefont {Shimoura}, \citenamefont {Sugimoto}, \citenamefont
  {Matsuta}, \citenamefont {Minamisono}, \citenamefont {Christie},
  \citenamefont {Olson},\ and\ \citenamefont
  {Wieman}}]{tanihata1992:11li-halo-density-correlation}\BibitemOpen
  \bibfield  {author} {\bibinfo {author} {\bibfnamefont {I.}~\bibnamefont
  {Tanihata}}, \bibinfo {author} {\bibfnamefont {T.}~\bibnamefont {Kobayashi}},
  \bibinfo {author} {\bibfnamefont {T.}~\bibnamefont {Suzuki}}, \bibinfo
  {author} {\bibfnamefont {K.}~\bibnamefont {Yoshida}}, \bibinfo {author}
  {\bibfnamefont {S.}~\bibnamefont {Shimoura}}, \bibinfo {author}
  {\bibfnamefont {K.}~\bibnamefont {Sugimoto}}, \bibinfo {author}
  {\bibfnamefont {K.}~\bibnamefont {Matsuta}}, \bibinfo {author} {\bibfnamefont
  {T.}~\bibnamefont {Minamisono}}, \bibinfo {author} {\bibfnamefont
  {W.}~\bibnamefont {Christie}}, \bibinfo {author} {\bibfnamefont
  {D.}~\bibnamefont {Olson}},\ and\ \bibinfo {author} {\bibfnamefont
  {H.}~\bibnamefont {Wieman}},\ }\href
  {https://doi.org/10.1016/0370-2693(92)90988-G} {\bibfield  {journal}
  {\bibinfo  {journal} {Phys. Lett. B}\ }\textbf {\bibinfo {volume} {287}},\
  \bibinfo {pages} {307} (\bibinfo {year} {1992})}\BibitemShut {NoStop}\bibitem [{\citenamefont {Simon}\ \emph {et~al.}(1999)\citenamefont {Simon},
  \citenamefont {Aleksandrov}, \citenamefont {Aumann}, \citenamefont
  {Axelsson}, \citenamefont {Baumann}, \citenamefont {Borge}, \citenamefont
  {Chulkov}, \citenamefont {Collatz}, \citenamefont {Cub}, \citenamefont
  {Dostal}, \citenamefont {Eberlein}, \citenamefont {Elze}, \citenamefont
  {Emling}, \citenamefont {Geissel}, \citenamefont {Gr\"unschloss},
  \citenamefont {Hellstr\"om}, \citenamefont {Holeczek}, \citenamefont
  {Holzmann}, \citenamefont {Jonson}, \citenamefont {Kratz}, \citenamefont
  {Kraus}, \citenamefont {Kulessa}, \citenamefont {Leifels}, \citenamefont
  {Leistenschneider}, \citenamefont {Leth}, \citenamefont {Mukha},
  \citenamefont {M\"unzenberg}, \citenamefont {Nickel}, \citenamefont
  {Nilsson}, \citenamefont {Nyman}, \citenamefont {Petersen}, \citenamefont
  {Pf\"utzner}, \citenamefont {Richter}, \citenamefont {Riisager},
  \citenamefont {Scheidenberger}, \citenamefont {Schrieder}, \citenamefont
  {Schwab}, \citenamefont {Smedberg}, \citenamefont {Stroth}, \citenamefont
  {Surowiec}, \citenamefont {Tengblad},\ and\ \citenamefont
  {Zhukov}}]{PhysRevLett.83.496}\BibitemOpen
  \bibfield  {author} {\bibinfo {author} {\bibfnamefont {H.}~\bibnamefont
  {Simon}}, \bibinfo {author} {\bibfnamefont {D.}~\bibnamefont {Aleksandrov}},
  \bibinfo {author} {\bibfnamefont {T.}~\bibnamefont {Aumann}}, \bibinfo
  {author} {\bibfnamefont {L.}~\bibnamefont {Axelsson}}, \bibinfo {author}
  {\bibfnamefont {T.}~\bibnamefont {Baumann}}, \bibinfo {author} {\bibfnamefont
  {M.~J.~G.}\ \bibnamefont {Borge}}, \bibinfo {author} {\bibfnamefont {L.~V.}\
  \bibnamefont {Chulkov}}, \bibinfo {author} {\bibfnamefont {R.}~\bibnamefont
  {Collatz}}, \bibinfo {author} {\bibfnamefont {J.}~\bibnamefont {Cub}},
  \bibinfo {author} {\bibfnamefont {W.}~\bibnamefont {Dostal}}, \bibinfo
  {author} {\bibfnamefont {B.}~\bibnamefont {Eberlein}}, \bibinfo {author}
  {\bibfnamefont {T.~W.}\ \bibnamefont {Elze}}, \bibinfo {author}
  {\bibfnamefont {H.}~\bibnamefont {Emling}}, \bibinfo {author} {\bibfnamefont
  {H.}~\bibnamefont {Geissel}}, \bibinfo {author} {\bibfnamefont
  {A.}~\bibnamefont {Gr\"unschloss}}, \bibinfo {author} {\bibfnamefont
  {M.}~\bibnamefont {Hellstr\"om}}, \bibinfo {author} {\bibfnamefont
  {J.}~\bibnamefont {Holeczek}}, \bibinfo {author} {\bibfnamefont
  {R.}~\bibnamefont {Holzmann}}, \bibinfo {author} {\bibfnamefont
  {B.}~\bibnamefont {Jonson}}, \bibinfo {author} {\bibfnamefont {J.~V.}\
  \bibnamefont {Kratz}}, \bibinfo {author} {\bibfnamefont {G.}~\bibnamefont
  {Kraus}}, \bibinfo {author} {\bibfnamefont {R.}~\bibnamefont {Kulessa}},
  \bibinfo {author} {\bibfnamefont {Y.}~\bibnamefont {Leifels}}, \bibinfo
  {author} {\bibfnamefont {A.}~\bibnamefont {Leistenschneider}}, \bibinfo
  {author} {\bibfnamefont {T.}~\bibnamefont {Leth}}, \bibinfo {author}
  {\bibfnamefont {I.}~\bibnamefont {Mukha}}, \bibinfo {author} {\bibfnamefont
  {G.}~\bibnamefont {M\"unzenberg}}, \bibinfo {author} {\bibfnamefont
  {F.}~\bibnamefont {Nickel}}, \bibinfo {author} {\bibfnamefont
  {T.}~\bibnamefont {Nilsson}}, \bibinfo {author} {\bibfnamefont
  {G.}~\bibnamefont {Nyman}}, \bibinfo {author} {\bibfnamefont
  {B.}~\bibnamefont {Petersen}}, \bibinfo {author} {\bibfnamefont
  {M.}~\bibnamefont {Pf\"utzner}}, \bibinfo {author} {\bibfnamefont
  {A.}~\bibnamefont {Richter}}, \bibinfo {author} {\bibfnamefont
  {K.}~\bibnamefont {Riisager}}, \bibinfo {author} {\bibfnamefont
  {C.}~\bibnamefont {Scheidenberger}}, \bibinfo {author} {\bibfnamefont
  {G.}~\bibnamefont {Schrieder}}, \bibinfo {author} {\bibfnamefont
  {W.}~\bibnamefont {Schwab}}, \bibinfo {author} {\bibfnamefont {M.~H.}\
  \bibnamefont {Smedberg}}, \bibinfo {author} {\bibfnamefont {J.}~\bibnamefont
  {Stroth}}, \bibinfo {author} {\bibfnamefont {A.}~\bibnamefont {Surowiec}},
  \bibinfo {author} {\bibfnamefont {O.}~\bibnamefont {Tengblad}},\ and\
  \bibinfo {author} {\bibfnamefont {M.~V.}\ \bibnamefont {Zhukov}},\ }\href
  {https://doi.org/10.1103/PhysRevLett.83.496} {\bibfield  {journal} {\bibinfo
  {journal} {Phys. Rev. Lett.}\ }\textbf {\bibinfo {volume} {83}},\ \bibinfo
  {pages} {496} (\bibinfo {year} {1999})}\BibitemShut {NoStop}\bibitem [{\citenamefont {Bagchi}\ \emph {et~al.}(2020)\citenamefont {Bagchi},
  \citenamefont {Kanungo}, \citenamefont {Tanaka}, \citenamefont {Geissel},
  \citenamefont {Doornenbal}, \citenamefont {Horiuchi}, \citenamefont {Hagen},
  \citenamefont {Suzuki}, \citenamefont {Tsunoda}, \citenamefont {Ahn},
  \citenamefont {Baba}, \citenamefont {Behr}, \citenamefont {Browne},
  \citenamefont {Chen}, \citenamefont {Cort\'es}, \citenamefont {Estrad\'e},
  \citenamefont {Fukuda}, \citenamefont {Holl}, \citenamefont {Itahashi},
  \citenamefont {Iwasa}, \citenamefont {Jansen}, \citenamefont {Jiang},
  \citenamefont {Kaur}, \citenamefont {Macchiavelli}, \citenamefont
  {Matsumoto}, \citenamefont {Momiyama}, \citenamefont {Murray}, \citenamefont
  {Nakamura}, \citenamefont {Novario}, \citenamefont {Ong}, \citenamefont
  {Otsuka}, \citenamefont {Papenbrock}, \citenamefont {Paschalis},
  \citenamefont {Prochazka}, \citenamefont {Scheidenberger}, \citenamefont
  {Schrock}, \citenamefont {Shimizu}, \citenamefont {Steppenbeck},
  \citenamefont {Sakurai}, \citenamefont {Suzuki}, \citenamefont {Suzuki},
  \citenamefont {Takechi}, \citenamefont {Takeda}, \citenamefont {Takeuchi},
  \citenamefont {Taniuchi}, \citenamefont {Wimmer},\ and\ \citenamefont
  {Yoshida}}]{PhysRevLett.124.222504}\BibitemOpen
  \bibfield  {author} {\bibinfo {author} {\bibfnamefont {S.}~\bibnamefont
  {Bagchi}}, \bibinfo {author} {\bibfnamefont {R.}~\bibnamefont {Kanungo}},
  \bibinfo {author} {\bibfnamefont {Y.~K.}\ \bibnamefont {Tanaka}}, \bibinfo
  {author} {\bibfnamefont {H.}~\bibnamefont {Geissel}}, \bibinfo {author}
  {\bibfnamefont {P.}~\bibnamefont {Doornenbal}}, \bibinfo {author}
  {\bibfnamefont {W.}~\bibnamefont {Horiuchi}}, \bibinfo {author}
  {\bibfnamefont {G.}~\bibnamefont {Hagen}}, \bibinfo {author} {\bibfnamefont
  {T.}~\bibnamefont {Suzuki}}, \bibinfo {author} {\bibfnamefont
  {N.}~\bibnamefont {Tsunoda}}, \bibinfo {author} {\bibfnamefont {D.~S.}\
  \bibnamefont {Ahn}}, \bibinfo {author} {\bibfnamefont {H.}~\bibnamefont
  {Baba}}, \bibinfo {author} {\bibfnamefont {K.}~\bibnamefont {Behr}}, \bibinfo
  {author} {\bibfnamefont {F.}~\bibnamefont {Browne}}, \bibinfo {author}
  {\bibfnamefont {S.}~\bibnamefont {Chen}}, \bibinfo {author} {\bibfnamefont
  {M.~L.}\ \bibnamefont {Cort\'es}}, \bibinfo {author} {\bibfnamefont
  {A.}~\bibnamefont {Estrad\'e}}, \bibinfo {author} {\bibfnamefont
  {N.}~\bibnamefont {Fukuda}}, \bibinfo {author} {\bibfnamefont
  {M.}~\bibnamefont {Holl}}, \bibinfo {author} {\bibfnamefont {K.}~\bibnamefont
  {Itahashi}}, \bibinfo {author} {\bibfnamefont {N.}~\bibnamefont {Iwasa}},
  \bibinfo {author} {\bibfnamefont {G.~R.}\ \bibnamefont {Jansen}}, \bibinfo
  {author} {\bibfnamefont {W.~G.}\ \bibnamefont {Jiang}}, \bibinfo {author}
  {\bibfnamefont {S.}~\bibnamefont {Kaur}}, \bibinfo {author} {\bibfnamefont
  {A.~O.}\ \bibnamefont {Macchiavelli}}, \bibinfo {author} {\bibfnamefont
  {S.~Y.}\ \bibnamefont {Matsumoto}}, \bibinfo {author} {\bibfnamefont
  {S.}~\bibnamefont {Momiyama}}, \bibinfo {author} {\bibfnamefont
  {I.}~\bibnamefont {Murray}}, \bibinfo {author} {\bibfnamefont
  {T.}~\bibnamefont {Nakamura}}, \bibinfo {author} {\bibfnamefont {S.~J.}\
  \bibnamefont {Novario}}, \bibinfo {author} {\bibfnamefont {H.~J.}\
  \bibnamefont {Ong}}, \bibinfo {author} {\bibfnamefont {T.}~\bibnamefont
  {Otsuka}}, \bibinfo {author} {\bibfnamefont {T.}~\bibnamefont {Papenbrock}},
  \bibinfo {author} {\bibfnamefont {S.}~\bibnamefont {Paschalis}}, \bibinfo
  {author} {\bibfnamefont {A.}~\bibnamefont {Prochazka}}, \bibinfo {author}
  {\bibfnamefont {C.}~\bibnamefont {Scheidenberger}}, \bibinfo {author}
  {\bibfnamefont {P.}~\bibnamefont {Schrock}}, \bibinfo {author} {\bibfnamefont
  {Y.}~\bibnamefont {Shimizu}}, \bibinfo {author} {\bibfnamefont
  {D.}~\bibnamefont {Steppenbeck}}, \bibinfo {author} {\bibfnamefont
  {H.}~\bibnamefont {Sakurai}}, \bibinfo {author} {\bibfnamefont
  {D.}~\bibnamefont {Suzuki}}, \bibinfo {author} {\bibfnamefont
  {H.}~\bibnamefont {Suzuki}}, \bibinfo {author} {\bibfnamefont
  {M.}~\bibnamefont {Takechi}}, \bibinfo {author} {\bibfnamefont
  {H.}~\bibnamefont {Takeda}}, \bibinfo {author} {\bibfnamefont
  {S.}~\bibnamefont {Takeuchi}}, \bibinfo {author} {\bibfnamefont
  {R.}~\bibnamefont {Taniuchi}}, \bibinfo {author} {\bibfnamefont
  {K.}~\bibnamefont {Wimmer}},\ and\ \bibinfo {author} {\bibfnamefont
  {K.}~\bibnamefont {Yoshida}},\ }\href
  {https://doi.org/10.1103/PhysRevLett.124.222504} {\bibfield  {journal}
  {\bibinfo  {journal} {Phys. Rev. Lett.}\ }\textbf {\bibinfo {volume} {124}},\
  \bibinfo {pages} {222504} (\bibinfo {year} {2020})}\BibitemShut {NoStop}\bibitem [{\citenamefont {Revel}\ \emph {et~al.}(2020)\citenamefont {Revel},
  \citenamefont {Sorlin}, \citenamefont {Marqu\'es}, \citenamefont {Kondo},
  \citenamefont {Kahlbow}, \citenamefont {Nakamura}, \citenamefont {Orr},
  \citenamefont {Nowacki}, \citenamefont {Tostevin}, \citenamefont {Yuan},
  \citenamefont {Achouri}, \citenamefont {Al~Falou}, \citenamefont {Atar},
  \citenamefont {Aumann}, \citenamefont {Baba}, \citenamefont {Boretzky},
  \citenamefont {Caesar}, \citenamefont {Calvet}, \citenamefont {Chae},
  \citenamefont {Chiga}, \citenamefont {Corsi}, \citenamefont {Crawford},
  \citenamefont {Delaunay}, \citenamefont {Delbart}, \citenamefont {Deshayes},
  \citenamefont {Dombr\'adi}, \citenamefont {Douma}, \citenamefont {Elekes},
  \citenamefont {Fallon}, \citenamefont {Ga\v{s}pari\'{c}}, \citenamefont
  {Gheller}, \citenamefont {Gibelin}, \citenamefont {Gillibert}, \citenamefont
  {Harakeh}, \citenamefont {He}, \citenamefont {Hirayama}, \citenamefont
  {Hoffman}, \citenamefont {Holl}, \citenamefont {Horvat}, \citenamefont
  {Horv\'ath}, \citenamefont {Hwang}, \citenamefont {Isobe}, \citenamefont
  {Kalantar-Nayestanaki}, \citenamefont {Kawase}, \citenamefont {Kim},
  \citenamefont {Kisamori}, \citenamefont {Kobayashi}, \citenamefont
  {K\"orper}, \citenamefont {Koyama}, \citenamefont {Kuti}, \citenamefont
  {Lapoux}, \citenamefont {Lindberg}, \citenamefont {Masuoka}, \citenamefont
  {Mayer}, \citenamefont {Miki}, \citenamefont {Murakami}, \citenamefont
  {Najafi}, \citenamefont {Nakano}, \citenamefont {Nakatsuka}, \citenamefont
  {Nilsson}, \citenamefont {Obertelli}, \citenamefont {de~Oliveira~Santos},
  \citenamefont {Otsu}, \citenamefont {Ozaki}, \citenamefont {Panin},
  \citenamefont {Paschalis}, \citenamefont {Rossi}, \citenamefont {Saito},
  \citenamefont {Saito}, \citenamefont {Sasano}, \citenamefont {Sato},
  \citenamefont {Satou}, \citenamefont {Scheit}, \citenamefont {Schindler},
  \citenamefont {Schrock}, \citenamefont {Shikata}, \citenamefont {Shimizu},
  \citenamefont {Simon}, \citenamefont {Sohler}, \citenamefont {Stuhl},
  \citenamefont {Takeuchi}, \citenamefont {Tanaka}, \citenamefont
  {Thoennessen}, \citenamefont {T\"ornqvist}, \citenamefont {Togano},
  \citenamefont {Tomai}, \citenamefont {Tscheuschner}, \citenamefont {Tsubota},
  \citenamefont {Uesaka}, \citenamefont {Yang}, \citenamefont {Yasuda},\ and\
  \citenamefont {Yoneda}}]{revel2020:28f-removal}\BibitemOpen
  \bibfield  {author} {\bibinfo {author} {\bibfnamefont {A.}~\bibnamefont
  {Revel}}, \bibinfo {author} {\bibfnamefont {O.}~\bibnamefont {Sorlin}},
  \bibinfo {author} {\bibfnamefont {F.~M.}\ \bibnamefont {Marqu\'es}}, \bibinfo
  {author} {\bibfnamefont {Y.}~\bibnamefont {Kondo}}, \bibinfo {author}
  {\bibfnamefont {J.}~\bibnamefont {Kahlbow}}, \bibinfo {author} {\bibfnamefont
  {T.}~\bibnamefont {Nakamura}}, \bibinfo {author} {\bibfnamefont {N.~A.}\
  \bibnamefont {Orr}}, \bibinfo {author} {\bibfnamefont {F.}~\bibnamefont
  {Nowacki}}, \bibinfo {author} {\bibfnamefont {J.~A.}\ \bibnamefont
  {Tostevin}}, \bibinfo {author} {\bibfnamefont {C.~X.}\ \bibnamefont {Yuan}},
  \bibinfo {author} {\bibfnamefont {N.~L.}\ \bibnamefont {Achouri}}, \bibinfo
  {author} {\bibfnamefont {H.}~\bibnamefont {Al~Falou}}, \bibinfo {author}
  {\bibfnamefont {L.}~\bibnamefont {Atar}}, \bibinfo {author} {\bibfnamefont
  {T.}~\bibnamefont {Aumann}}, \bibinfo {author} {\bibfnamefont
  {H.}~\bibnamefont {Baba}}, \bibinfo {author} {\bibfnamefont {K.}~\bibnamefont
  {Boretzky}}, \bibinfo {author} {\bibfnamefont {C.}~\bibnamefont {Caesar}},
  \bibinfo {author} {\bibfnamefont {D.}~\bibnamefont {Calvet}}, \bibinfo
  {author} {\bibfnamefont {H.}~\bibnamefont {Chae}}, \bibinfo {author}
  {\bibfnamefont {N.}~\bibnamefont {Chiga}}, \bibinfo {author} {\bibfnamefont
  {A.}~\bibnamefont {Corsi}}, \bibinfo {author} {\bibfnamefont {H.~L.}\
  \bibnamefont {Crawford}}, \bibinfo {author} {\bibfnamefont {F.}~\bibnamefont
  {Delaunay}}, \bibinfo {author} {\bibfnamefont {A.}~\bibnamefont {Delbart}},
  \bibinfo {author} {\bibfnamefont {Q.}~\bibnamefont {Deshayes}}, \bibinfo
  {author} {\bibfnamefont {Z.}~\bibnamefont {Dombr\'adi}}, \bibinfo {author}
  {\bibfnamefont {C.~A.}\ \bibnamefont {Douma}}, \bibinfo {author}
  {\bibfnamefont {Z.}~\bibnamefont {Elekes}}, \bibinfo {author} {\bibfnamefont
  {P.}~\bibnamefont {Fallon}}, \bibinfo {author} {\bibfnamefont
  {I.}~\bibnamefont {Ga\v{s}pari\'{c}}}, \bibinfo {author} {\bibfnamefont
  {J.-M.}\ \bibnamefont {Gheller}}, \bibinfo {author} {\bibfnamefont
  {J.}~\bibnamefont {Gibelin}}, \bibinfo {author} {\bibfnamefont
  {A.}~\bibnamefont {Gillibert}}, \bibinfo {author} {\bibfnamefont {M.~N.}\
  \bibnamefont {Harakeh}}, \bibinfo {author} {\bibfnamefont {W.}~\bibnamefont
  {He}}, \bibinfo {author} {\bibfnamefont {A.}~\bibnamefont {Hirayama}},
  \bibinfo {author} {\bibfnamefont {C.~R.}\ \bibnamefont {Hoffman}}, \bibinfo
  {author} {\bibfnamefont {M.}~\bibnamefont {Holl}}, \bibinfo {author}
  {\bibfnamefont {A.}~\bibnamefont {Horvat}}, \bibinfo {author} {\bibfnamefont
  {A.}~\bibnamefont {Horv\'ath}}, \bibinfo {author} {\bibfnamefont {J.~W.}\
  \bibnamefont {Hwang}}, \bibinfo {author} {\bibfnamefont {T.}~\bibnamefont
  {Isobe}}, \bibinfo {author} {\bibfnamefont {N.}~\bibnamefont
  {Kalantar-Nayestanaki}}, \bibinfo {author} {\bibfnamefont {S.}~\bibnamefont
  {Kawase}}, \bibinfo {author} {\bibfnamefont {S.}~\bibnamefont {Kim}},
  \bibinfo {author} {\bibfnamefont {K.}~\bibnamefont {Kisamori}}, \bibinfo
  {author} {\bibfnamefont {T.}~\bibnamefont {Kobayashi}}, \bibinfo {author}
  {\bibfnamefont {D.}~\bibnamefont {K\"orper}}, \bibinfo {author}
  {\bibfnamefont {S.}~\bibnamefont {Koyama}}, \bibinfo {author} {\bibfnamefont
  {I.}~\bibnamefont {Kuti}}, \bibinfo {author} {\bibfnamefont {V.}~\bibnamefont
  {Lapoux}}, \bibinfo {author} {\bibfnamefont {S.}~\bibnamefont {Lindberg}},
  \bibinfo {author} {\bibfnamefont {S.}~\bibnamefont {Masuoka}}, \bibinfo
  {author} {\bibfnamefont {J.}~\bibnamefont {Mayer}}, \bibinfo {author}
  {\bibfnamefont {K.}~\bibnamefont {Miki}}, \bibinfo {author} {\bibfnamefont
  {T.}~\bibnamefont {Murakami}}, \bibinfo {author} {\bibfnamefont
  {M.}~\bibnamefont {Najafi}}, \bibinfo {author} {\bibfnamefont
  {K.}~\bibnamefont {Nakano}}, \bibinfo {author} {\bibfnamefont
  {N.}~\bibnamefont {Nakatsuka}}, \bibinfo {author} {\bibfnamefont
  {T.}~\bibnamefont {Nilsson}}, \bibinfo {author} {\bibfnamefont
  {A.}~\bibnamefont {Obertelli}}, \bibinfo {author} {\bibfnamefont
  {F.}~\bibnamefont {de~Oliveira~Santos}}, \bibinfo {author} {\bibfnamefont
  {H.}~\bibnamefont {Otsu}}, \bibinfo {author} {\bibfnamefont {T.}~\bibnamefont
  {Ozaki}}, \bibinfo {author} {\bibfnamefont {V.}~\bibnamefont {Panin}},
  \bibinfo {author} {\bibfnamefont {S.}~\bibnamefont {Paschalis}}, \bibinfo
  {author} {\bibfnamefont {D.}~\bibnamefont {Rossi}}, \bibinfo {author}
  {\bibfnamefont {A.~T.}\ \bibnamefont {Saito}}, \bibinfo {author}
  {\bibfnamefont {T.}~\bibnamefont {Saito}}, \bibinfo {author} {\bibfnamefont
  {M.}~\bibnamefont {Sasano}}, \bibinfo {author} {\bibfnamefont
  {H.}~\bibnamefont {Sato}}, \bibinfo {author} {\bibfnamefont {Y.}~\bibnamefont
  {Satou}}, \bibinfo {author} {\bibfnamefont {H.}~\bibnamefont {Scheit}},
  \bibinfo {author} {\bibfnamefont {F.}~\bibnamefont {Schindler}}, \bibinfo
  {author} {\bibfnamefont {P.}~\bibnamefont {Schrock}}, \bibinfo {author}
  {\bibfnamefont {M.}~\bibnamefont {Shikata}}, \bibinfo {author} {\bibfnamefont
  {Y.}~\bibnamefont {Shimizu}}, \bibinfo {author} {\bibfnamefont
  {H.}~\bibnamefont {Simon}}, \bibinfo {author} {\bibfnamefont
  {D.}~\bibnamefont {Sohler}}, \bibinfo {author} {\bibfnamefont
  {L.}~\bibnamefont {Stuhl}}, \bibinfo {author} {\bibfnamefont
  {S.}~\bibnamefont {Takeuchi}}, \bibinfo {author} {\bibfnamefont
  {M.}~\bibnamefont {Tanaka}}, \bibinfo {author} {\bibfnamefont
  {M.}~\bibnamefont {Thoennessen}}, \bibinfo {author} {\bibfnamefont
  {H.}~\bibnamefont {T\"ornqvist}}, \bibinfo {author} {\bibfnamefont
  {Y.}~\bibnamefont {Togano}}, \bibinfo {author} {\bibfnamefont
  {T.}~\bibnamefont {Tomai}}, \bibinfo {author} {\bibfnamefont
  {J.}~\bibnamefont {Tscheuschner}}, \bibinfo {author} {\bibfnamefont
  {J.}~\bibnamefont {Tsubota}}, \bibinfo {author} {\bibfnamefont
  {T.}~\bibnamefont {Uesaka}}, \bibinfo {author} {\bibfnamefont
  {Z.}~\bibnamefont {Yang}}, \bibinfo {author} {\bibfnamefont {M.}~\bibnamefont
  {Yasuda}},\ and\ \bibinfo {author} {\bibfnamefont {K.}~\bibnamefont
  {Yoneda}},\ }\href {https://doi.org/10.1103/PhysRevLett.124.152502}
  {\bibfield  {journal} {\bibinfo  {journal} {Phys. Rev. Lett.}\ }\textbf
  {\bibinfo {volume} {124}},\ \bibinfo {pages} {152502} (\bibinfo {year}
  {2020})}\BibitemShut {NoStop}\bibitem [{\citenamefont {Wang}\ \emph {et~al.}(2023)\citenamefont {Wang},
  \citenamefont {Yasuda}, \citenamefont {Kondo}, \citenamefont {Nakamura},
  \citenamefont {Tostevin}, \citenamefont {Ogata}, \citenamefont {Otsuka},
  \citenamefont {Poves}, \citenamefont {Shimizu}, \citenamefont {Yoshida},
  \citenamefont {Achouri}, \citenamefont {{Al Falou}}, \citenamefont {Atar},
  \citenamefont {Aumann}, \citenamefont {Baba}, \citenamefont {Boretzky},
  \citenamefont {Caesar}, \citenamefont {Calvet}, \citenamefont {Chae},
  \citenamefont {Chiga}, \citenamefont {Corsi}, \citenamefont {Crawford},
  \citenamefont {Delaunay}, \citenamefont {Delbart}, \citenamefont {Deshayes},
  \citenamefont {Dombr\'{a}di}, \citenamefont {Douma}, \citenamefont {Elekes},
  \citenamefont {Fallon}, \citenamefont {Ga\v{s}pari\'{c}}, \citenamefont
  {Gheller}, \citenamefont {Gibelin}, \citenamefont {Gillibert}, \citenamefont
  {Harakeh}, \citenamefont {Hirayama}, \citenamefont {Hoffman}, \citenamefont
  {Holl}, \citenamefont {Horvat}, \citenamefont {Horv'{a}th}, \citenamefont
  {Hwang}, \citenamefont {Isobe}, \citenamefont {Kahlbow}, \citenamefont
  {Kalantar-Nayestanaki}, \citenamefont {Kawase}, \citenamefont {Kim},
  \citenamefont {Kisamori}, \citenamefont {Kobayashi}, \citenamefont
  {K\"{o}rper}, \citenamefont {Koyama}, \citenamefont {Kuti}, \citenamefont
  {Lapoux}, \citenamefont {Lindberg}, \citenamefont {Marqu\'{e}s},
  \citenamefont {Masuoka}, \citenamefont {Mayer}, \citenamefont {Miki},
  \citenamefont {Murakami}, \citenamefont {Najafi}, \citenamefont {Nakano},
  \citenamefont {Nakatsuka}, \citenamefont {Nilsson}, \citenamefont
  {Obertelli}, \citenamefont {Orr}, \citenamefont {Otsu}, \citenamefont
  {Ozaki}, \citenamefont {Panin}, \citenamefont {Paschalis}, \citenamefont
  {Revel}, \citenamefont {Rossi}, \citenamefont {Saito}, \citenamefont {Saito},
  \citenamefont {Sasano}, \citenamefont {Sato}, \citenamefont {Satou},
  \citenamefont {Scheit}, \citenamefont {Schindler}, \citenamefont {Schrock},
  \citenamefont {Shikata}, \citenamefont {Shimizu}, \citenamefont {Simon},
  \citenamefont {Sohler}, \citenamefont {Sorlin}, \citenamefont {Stuhl},
  \citenamefont {Takeuchi}, \citenamefont {Tanaka}, \citenamefont
  {Thoennessen}, \citenamefont {T\"{o}rnqvist}, \citenamefont {Togano},
  \citenamefont {Tomai}, \citenamefont {Tscheuschner}, \citenamefont {Tsubota},
  \citenamefont {Uesaka}, \citenamefont {Yang},\ and\ \citenamefont
  {Yoneda}}]{wang2023intruder}\BibitemOpen
  \bibfield  {author} {\bibinfo {author} {\bibfnamefont {H.}~\bibnamefont
  {Wang}}, \bibinfo {author} {\bibfnamefont {M.}~\bibnamefont {Yasuda}},
  \bibinfo {author} {\bibfnamefont {Y.}~\bibnamefont {Kondo}}, \bibinfo
  {author} {\bibfnamefont {T.}~\bibnamefont {Nakamura}}, \bibinfo {author}
  {\bibfnamefont {J.~A.}\ \bibnamefont {Tostevin}}, \bibinfo {author}
  {\bibfnamefont {K.}~\bibnamefont {Ogata}}, \bibinfo {author} {\bibfnamefont
  {T.}~\bibnamefont {Otsuka}}, \bibinfo {author} {\bibfnamefont
  {A.}~\bibnamefont {Poves}}, \bibinfo {author} {\bibfnamefont
  {N.}~\bibnamefont {Shimizu}}, \bibinfo {author} {\bibfnamefont
  {K.}~\bibnamefont {Yoshida}}, \bibinfo {author} {\bibfnamefont {N.~L.}\
  \bibnamefont {Achouri}}, \bibinfo {author} {\bibfnamefont {H.}~\bibnamefont
  {{Al Falou}}}, \bibinfo {author} {\bibfnamefont {L.}~\bibnamefont {Atar}},
  \bibinfo {author} {\bibfnamefont {T.}~\bibnamefont {Aumann}}, \bibinfo
  {author} {\bibfnamefont {H.}~\bibnamefont {Baba}}, \bibinfo {author}
  {\bibfnamefont {K.}~\bibnamefont {Boretzky}}, \bibinfo {author}
  {\bibfnamefont {C.}~\bibnamefont {Caesar}}, \bibinfo {author} {\bibfnamefont
  {D.}~\bibnamefont {Calvet}}, \bibinfo {author} {\bibfnamefont
  {H.}~\bibnamefont {Chae}}, \bibinfo {author} {\bibfnamefont {N.}~\bibnamefont
  {Chiga}}, \bibinfo {author} {\bibfnamefont {A.}~\bibnamefont {Corsi}},
  \bibinfo {author} {\bibfnamefont {H.~L.}\ \bibnamefont {Crawford}}, \bibinfo
  {author} {\bibfnamefont {F.}~\bibnamefont {Delaunay}}, \bibinfo {author}
  {\bibfnamefont {A.}~\bibnamefont {Delbart}}, \bibinfo {author} {\bibfnamefont
  {Q.}~\bibnamefont {Deshayes}}, \bibinfo {author} {\bibfnamefont
  {Z.}~\bibnamefont {Dombr\'{a}di}}, \bibinfo {author} {\bibfnamefont
  {C.}~\bibnamefont {Douma}}, \bibinfo {author} {\bibfnamefont
  {Z.}~\bibnamefont {Elekes}}, \bibinfo {author} {\bibfnamefont
  {P.}~\bibnamefont {Fallon}}, \bibinfo {author} {\bibfnamefont
  {I.}~\bibnamefont {Ga\v{s}pari\'{c}}}, \bibinfo {author} {\bibfnamefont
  {J.-M.}\ \bibnamefont {Gheller}}, \bibinfo {author} {\bibfnamefont
  {J.}~\bibnamefont {Gibelin}}, \bibinfo {author} {\bibfnamefont
  {A.}~\bibnamefont {Gillibert}}, \bibinfo {author} {\bibfnamefont {M.~N.}\
  \bibnamefont {Harakeh}}, \bibinfo {author} {\bibfnamefont {A.}~\bibnamefont
  {Hirayama}}, \bibinfo {author} {\bibfnamefont {C.~R.}\ \bibnamefont
  {Hoffman}}, \bibinfo {author} {\bibfnamefont {M.}~\bibnamefont {Holl}},
  \bibinfo {author} {\bibfnamefont {A.}~\bibnamefont {Horvat}}, \bibinfo
  {author} {\bibfnamefont {A.}~\bibnamefont {Horv'{a}th}}, \bibinfo {author}
  {\bibfnamefont {J.~W.}\ \bibnamefont {Hwang}}, \bibinfo {author}
  {\bibfnamefont {T.}~\bibnamefont {Isobe}}, \bibinfo {author} {\bibfnamefont
  {J.}~\bibnamefont {Kahlbow}}, \bibinfo {author} {\bibfnamefont
  {N.}~\bibnamefont {Kalantar-Nayestanaki}}, \bibinfo {author} {\bibfnamefont
  {S.}~\bibnamefont {Kawase}}, \bibinfo {author} {\bibfnamefont
  {S.}~\bibnamefont {Kim}}, \bibinfo {author} {\bibfnamefont {K.}~\bibnamefont
  {Kisamori}}, \bibinfo {author} {\bibfnamefont {T.}~\bibnamefont {Kobayashi}},
  \bibinfo {author} {\bibfnamefont {D.}~\bibnamefont {K\"{o}rper}}, \bibinfo
  {author} {\bibfnamefont {S.}~\bibnamefont {Koyama}}, \bibinfo {author}
  {\bibfnamefont {I.}~\bibnamefont {Kuti}}, \bibinfo {author} {\bibfnamefont
  {V.}~\bibnamefont {Lapoux}}, \bibinfo {author} {\bibfnamefont
  {S.}~\bibnamefont {Lindberg}}, \bibinfo {author} {\bibfnamefont {F.~M.}\
  \bibnamefont {Marqu\'{e}s}}, \bibinfo {author} {\bibfnamefont
  {S.}~\bibnamefont {Masuoka}}, \bibinfo {author} {\bibfnamefont
  {J.}~\bibnamefont {Mayer}}, \bibinfo {author} {\bibfnamefont
  {K.}~\bibnamefont {Miki}}, \bibinfo {author} {\bibfnamefont {T.}~\bibnamefont
  {Murakami}}, \bibinfo {author} {\bibfnamefont {M.~A.}\ \bibnamefont
  {Najafi}}, \bibinfo {author} {\bibfnamefont {K.}~\bibnamefont {Nakano}},
  \bibinfo {author} {\bibfnamefont {N.}~\bibnamefont {Nakatsuka}}, \bibinfo
  {author} {\bibfnamefont {T.}~\bibnamefont {Nilsson}}, \bibinfo {author}
  {\bibfnamefont {A.}~\bibnamefont {Obertelli}}, \bibinfo {author}
  {\bibfnamefont {N.~A.}\ \bibnamefont {Orr}}, \bibinfo {author} {\bibfnamefont
  {H.}~\bibnamefont {Otsu}}, \bibinfo {author} {\bibfnamefont {T.}~\bibnamefont
  {Ozaki}}, \bibinfo {author} {\bibfnamefont {V.}~\bibnamefont {Panin}},
  \bibinfo {author} {\bibfnamefont {S.}~\bibnamefont {Paschalis}}, \bibinfo
  {author} {\bibfnamefont {A.}~\bibnamefont {Revel}}, \bibinfo {author}
  {\bibfnamefont {D.}~\bibnamefont {Rossi}}, \bibinfo {author} {\bibfnamefont
  {A.~T.}\ \bibnamefont {Saito}}, \bibinfo {author} {\bibfnamefont
  {T.}~\bibnamefont {Saito}}, \bibinfo {author} {\bibfnamefont
  {M.}~\bibnamefont {Sasano}}, \bibinfo {author} {\bibfnamefont
  {H.}~\bibnamefont {Sato}}, \bibinfo {author} {\bibfnamefont {Y.}~\bibnamefont
  {Satou}}, \bibinfo {author} {\bibfnamefont {H.}~\bibnamefont {Scheit}},
  \bibinfo {author} {\bibfnamefont {F.}~\bibnamefont {Schindler}}, \bibinfo
  {author} {\bibfnamefont {P.}~\bibnamefont {Schrock}}, \bibinfo {author}
  {\bibfnamefont {M.}~\bibnamefont {Shikata}}, \bibinfo {author} {\bibfnamefont
  {Y.}~\bibnamefont {Shimizu}}, \bibinfo {author} {\bibfnamefont
  {H.}~\bibnamefont {Simon}}, \bibinfo {author} {\bibfnamefont
  {D.}~\bibnamefont {Sohler}}, \bibinfo {author} {\bibfnamefont
  {O.}~\bibnamefont {Sorlin}}, \bibinfo {author} {\bibfnamefont
  {L.}~\bibnamefont {Stuhl}}, \bibinfo {author} {\bibfnamefont
  {S.}~\bibnamefont {Takeuchi}}, \bibinfo {author} {\bibfnamefont
  {M.}~\bibnamefont {Tanaka}}, \bibinfo {author} {\bibfnamefont
  {M.}~\bibnamefont {Thoennessen}}, \bibinfo {author} {\bibfnamefont
  {H.}~\bibnamefont {T\"{o}rnqvist}}, \bibinfo {author} {\bibfnamefont
  {Y.}~\bibnamefont {Togano}}, \bibinfo {author} {\bibfnamefont
  {T.}~\bibnamefont {Tomai}}, \bibinfo {author} {\bibfnamefont
  {J.}~\bibnamefont {Tscheuschner}}, \bibinfo {author} {\bibfnamefont
  {J.}~\bibnamefont {Tsubota}}, \bibinfo {author} {\bibfnamefont
  {T.}~\bibnamefont {Uesaka}}, \bibinfo {author} {\bibfnamefont
  {Z.}~\bibnamefont {Yang}},\ and\ \bibinfo {author} {\bibfnamefont
  {K.}~\bibnamefont {Yoneda}},\ }\href
  {https://doi.org/10.1016/j.physletb.2023.138038} {\bibfield  {journal}
  {\bibinfo  {journal} {Phys. Lett. B}\ }\textbf {\bibinfo {volume} {843}},\
  \bibinfo {pages} {138038} (\bibinfo {year} {2023})}\BibitemShut {NoStop}\bibitem [{\citenamefont {Macchiavelli}\ \emph {et~al.}(2017)\citenamefont
  {Macchiavelli}, \citenamefont {Crawford}, \citenamefont {Fallon},
  \citenamefont {Campbell}, \citenamefont {Clark}, \citenamefont {Cromaz},
  \citenamefont {Jones}, \citenamefont {Lee},\ and\ \citenamefont
  {Salathe}}]{macchiavelli2017:29f-nilsson-rotation-aligned}\BibitemOpen
  \bibfield  {author} {\bibinfo {author} {\bibfnamefont {A.~O.}\ \bibnamefont
  {Macchiavelli}}, \bibinfo {author} {\bibfnamefont {H.~L.}\ \bibnamefont
  {Crawford}}, \bibinfo {author} {\bibfnamefont {P.}~\bibnamefont {Fallon}},
  \bibinfo {author} {\bibfnamefont {C.~M.}\ \bibnamefont {Campbell}}, \bibinfo
  {author} {\bibfnamefont {R.~M.}\ \bibnamefont {Clark}}, \bibinfo {author}
  {\bibfnamefont {M.}~\bibnamefont {Cromaz}}, \bibinfo {author} {\bibfnamefont
  {M.~D.}\ \bibnamefont {Jones}}, \bibinfo {author} {\bibfnamefont {I.~Y.}\
  \bibnamefont {Lee}},\ and\ \bibinfo {author} {\bibfnamefont {M.}~\bibnamefont
  {Salathe}},\ }\href {https://doi.org/10.1016/j.physletb.2017.10.041}
  {\bibfield  {journal} {\bibinfo  {journal} {Phys. Lett. B}\ }\textbf
  {\bibinfo {volume} {775}},\ \bibinfo {pages} {160} (\bibinfo {year}
  {2017})}\BibitemShut {NoStop}\bibitem [{\citenamefont {Luo}\ \emph {et~al.}(2021)\citenamefont {Luo},
  \citenamefont {Fossez}, \citenamefont {Liu},\ and\ \citenamefont
  {Guo}}]{PhysRevC.104.014307}\BibitemOpen
  \bibfield  {author} {\bibinfo {author} {\bibfnamefont {Y.-X.}\ \bibnamefont
  {Luo}}, \bibinfo {author} {\bibfnamefont {K.}~\bibnamefont {Fossez}},
  \bibinfo {author} {\bibfnamefont {Q.}~\bibnamefont {Liu}},\ and\ \bibinfo
  {author} {\bibfnamefont {J.-Y.}\ \bibnamefont {Guo}},\ }\href
  {https://doi.org/10.1103/PhysRevC.104.014307} {\bibfield  {journal} {\bibinfo
   {journal} {Phys. Rev. C}\ }\textbf {\bibinfo {volume} {104}},\ \bibinfo
  {pages} {014307} (\bibinfo {year} {2021})}\BibitemShut {NoStop}\bibitem [{\citenamefont {Fortunato}\ \emph {et~al.}(2020)\citenamefont
  {Fortunato}, \citenamefont {Casal}, \citenamefont {Horiuchi}, \citenamefont
  {Singh},\ and\ \citenamefont {Vitturi}}]{fortunato202029f}\BibitemOpen
  \bibfield  {author} {\bibinfo {author} {\bibfnamefont {L.}~\bibnamefont
  {Fortunato}}, \bibinfo {author} {\bibfnamefont {J.}~\bibnamefont {Casal}},
  \bibinfo {author} {\bibfnamefont {W.}~\bibnamefont {Horiuchi}}, \bibinfo
  {author} {\bibfnamefont {J.}~\bibnamefont {Singh}},\ and\ \bibinfo {author}
  {\bibfnamefont {A.}~\bibnamefont {Vitturi}},\ }\href
  {https://doi.org/10.1038/s42005-020-00402-5} {\bibfield  {journal} {\bibinfo
  {journal} {Commun. Phys.}\ }\textbf {\bibinfo {volume} {3}},\ \bibinfo
  {pages} {132} (\bibinfo {year} {2020})}\BibitemShut {NoStop}\bibitem [{\citenamefont {Johnson}\ \emph {et~al.}(2018)\citenamefont
  {Johnson}, \citenamefont {Ormand}, \citenamefont {McElvain},\ and\
  \citenamefont {Shan}}]{johnson2018:bigstick}\BibitemOpen
  \bibfield  {author} {\bibinfo {author} {\bibfnamefont {C.~W.}\ \bibnamefont
  {Johnson}}, \bibinfo {author} {\bibfnamefont {W.~E.}\ \bibnamefont {Ormand}},
  \bibinfo {author} {\bibfnamefont {K.~S.}\ \bibnamefont {McElvain}},\ and\
  \bibinfo {author} {\bibfnamefont {H.}~\bibnamefont {Shan}},\ }\href@noop {}
  {\bibinfo {title} {\texttt{BIGSTICK}: A flexible configuration-interaction
  shell-model code}} (\bibinfo {year} {2018}),\ \Eprint
  {https://arxiv.org/abs/1801.08432} {arXiv:1801.08432 [physics.comp-ph]}
  \BibitemShut {NoStop}\bibitem [{\citenamefont {Maris}\ \emph {et~al.}(2010)\citenamefont {Maris},
  \citenamefont {Sosonkina}, \citenamefont {Vary}, \citenamefont {Ng},\ and\
  \citenamefont {Yang}}]{maris2010:ncsm-mfdn-iccs10}\BibitemOpen
  \bibfield  {author} {\bibinfo {author} {\bibfnamefont {P.}~\bibnamefont
  {Maris}}, \bibinfo {author} {\bibfnamefont {M.}~\bibnamefont {Sosonkina}},
  \bibinfo {author} {\bibfnamefont {J.~P.}\ \bibnamefont {Vary}}, \bibinfo
  {author} {\bibfnamefont {E.}~\bibnamefont {Ng}},\ and\ \bibinfo {author}
  {\bibfnamefont {C.}~\bibnamefont {Yang}},\ }\href
  {https://doi.org/10.1016/j.procs.2010.04.012} {\bibfield  {journal} {\bibinfo
   {journal} {Procedia Comput. Sci.}\ }\textbf {\bibinfo {volume} {1}},\
  \bibinfo {pages} {97} (\bibinfo {year} {2010})}\BibitemShut {NoStop}\bibitem [{\citenamefont {Shao}\ \emph {et~al.}(2018)\citenamefont {Shao},
  \citenamefont {Aktulga}, \citenamefont {Yang}, \citenamefont {Ng},
  \citenamefont {Maris},\ and\ \citenamefont
  {Vary}}]{shao2018:ncci-preconditioned}\BibitemOpen
  \bibfield  {author} {\bibinfo {author} {\bibfnamefont {M.}~\bibnamefont
  {Shao}}, \bibinfo {author} {\bibfnamefont {H.~M.}\ \bibnamefont {Aktulga}},
  \bibinfo {author} {\bibfnamefont {C.}~\bibnamefont {Yang}}, \bibinfo {author}
  {\bibfnamefont {E.~G.}\ \bibnamefont {Ng}}, \bibinfo {author} {\bibfnamefont
  {P.}~\bibnamefont {Maris}},\ and\ \bibinfo {author} {\bibfnamefont {J.~P.}\
  \bibnamefont {Vary}},\ }\href {https://doi.org/10.1016/j.cpc.2017.09.004}
  {\bibfield  {journal} {\bibinfo  {journal} {Comput. Phys. Commun.}\ }\textbf
  {\bibinfo {volume} {222}},\ \bibinfo {pages} {1} (\bibinfo {year}
  {2018})}\BibitemShut {NoStop}\bibitem [{\citenamefont {Fasano}\ and\ \citenamefont
  {Maris}(2025)}]{code-mfdn-transitions}\BibitemOpen
  \bibfield  {author} {\bibinfo {author} {\bibfnamefont {P.~J.}\ \bibnamefont
  {Fasano}}\ and\ \bibinfo {author} {\bibfnamefont {P.}~\bibnamefont {Maris}},\
  }\href {https://doi.org/10.5281/zenodo.18013361} {\bibinfo {title}
  {\textup{computer code \texttt{mfdn-transitions}}}} (\bibinfo {year}
  {2025})\BibitemShut {NoStop}\bibitem [{\citenamefont {Entem}\ and\ \citenamefont
  {Machleidt}(2003)}]{PhysRevC.68.041001}\BibitemOpen
  \bibfield  {author} {\bibinfo {author} {\bibfnamefont {D.~R.}\ \bibnamefont
  {Entem}}\ and\ \bibinfo {author} {\bibfnamefont {R.}~\bibnamefont
  {Machleidt}},\ }\href {https://doi.org/10.1103/PhysRevC.68.041001} {\bibfield
   {journal} {\bibinfo  {journal} {Phys. Rev. C}\ }\textbf {\bibinfo {volume}
  {68}},\ \bibinfo {pages} {041001} (\bibinfo {year} {2003})}\BibitemShut
  {NoStop}\bibitem [{\citenamefont {Shirokov}\ \emph {et~al.}(2016)\citenamefont
  {Shirokov}, \citenamefont {Shin}, \citenamefont {Kim}, \citenamefont
  {Sosonkina}, \citenamefont {Maris},\ and\ \citenamefont
  {Vary}}]{shirokov2016:nn-daejeon16}\BibitemOpen
  \bibfield  {author} {\bibinfo {author} {\bibfnamefont {A.~M.}\ \bibnamefont
  {Shirokov}}, \bibinfo {author} {\bibfnamefont {I.~J.}\ \bibnamefont {Shin}},
  \bibinfo {author} {\bibfnamefont {Y.}~\bibnamefont {Kim}}, \bibinfo {author}
  {\bibfnamefont {M.}~\bibnamefont {Sosonkina}}, \bibinfo {author}
  {\bibfnamefont {P.}~\bibnamefont {Maris}},\ and\ \bibinfo {author}
  {\bibfnamefont {J.~P.}\ \bibnamefont {Vary}},\ }\href
  {https://doi.org/10.1016/j.physletb.2016.08.006} {\bibfield  {journal}
  {\bibinfo  {journal} {Phys. Lett. B}\ }\textbf {\bibinfo {volume} {761}},\
  \bibinfo {pages} {87} (\bibinfo {year} {2016})}\BibitemShut {NoStop}\bibitem [{\citenamefont {Whitehead}(1980)}]{whitehead1980:lanczos}\BibitemOpen
  \bibfield  {author} {\bibinfo {author} {\bibfnamefont {R.~R.}\ \bibnamefont
  {Whitehead}},\ }in\ \href@noop {} {\emph {\bibinfo {booktitle} {Theory and
  Applications of Moment Methods in Many-Fermion Systems}}},\ \bibinfo {editor}
  {edited by\ \bibinfo {editor} {\bibfnamefont {B.~J.}\ \bibnamefont {Dalton}},
  \bibinfo {editor} {\bibfnamefont {S.~M.}\ \bibnamefont {Grimes}}, \bibinfo
  {editor} {\bibfnamefont {J.~P.}\ \bibnamefont {Vary}},\ and\ \bibinfo
  {editor} {\bibfnamefont {S.~A.}\ \bibnamefont {Williams}}}\ (\bibinfo
  {publisher} {Plenum},\ \bibinfo {address} {New York},\ \bibinfo {year}
  {1980})\ p.\ \bibinfo {pages} {235}\BibitemShut {NoStop}\bibitem [{\citenamefont {Gueorguiev}\ \emph {et~al.}(2000)\citenamefont
  {Gueorguiev}, \citenamefont {Draayer},\ and\ \citenamefont
  {Johnson}}]{gueorguiev2000:fp-su3-breaking}\BibitemOpen
  \bibfield  {author} {\bibinfo {author} {\bibfnamefont {V.~G.}\ \bibnamefont
  {Gueorguiev}}, \bibinfo {author} {\bibfnamefont {J.~P.}\ \bibnamefont
  {Draayer}},\ and\ \bibinfo {author} {\bibfnamefont {C.~W.}\ \bibnamefont
  {Johnson}},\ }\href {https://doi.org/10.1103/PhysRevC.63.014318} {\bibfield
  {journal} {\bibinfo  {journal} {Phys. Rev. C}\ }\textbf {\bibinfo {volume}
  {63}},\ \bibinfo {pages} {014318} (\bibinfo {year} {2000})}\BibitemShut
  {NoStop}\bibitem [{\citenamefont {Johnson}(2015)}]{PhysRevC.91.034313}\BibitemOpen
  \bibfield  {author} {\bibinfo {author} {\bibfnamefont {C.~W.}\ \bibnamefont
  {Johnson}},\ }\href {https://doi.org/10.1103/PhysRevC.91.034313} {\bibfield
  {journal} {\bibinfo  {journal} {Phys. Rev. C}\ }\textbf {\bibinfo {volume}
  {91}},\ \bibinfo {pages} {034313} (\bibinfo {year} {2015})}\BibitemShut
  {NoStop}\bibitem [{\citenamefont {Stone}(2016)}]{stone2016:e2-moments}\BibitemOpen
  \bibfield  {author} {\bibinfo {author} {\bibfnamefont {N.~J.}\ \bibnamefont
  {Stone}},\ }\href {https://doi.org/10.1016/j.adt.2015.12.002} {\bibfield
  {journal} {\bibinfo  {journal} {At. Data Nucl. Data Tables}\ }\textbf
  {\bibinfo {volume} {111--112}},\ \bibinfo {pages} {1} (\bibinfo {year}
  {2016})}\BibitemShut {NoStop}\bibitem [{\citenamefont {Angeli}\ and\ \citenamefont
  {Marinova}(2013)}]{angeli2013:charge-radii}\BibitemOpen
  \bibfield  {author} {\bibinfo {author} {\bibfnamefont {I.}~\bibnamefont
  {Angeli}}\ and\ \bibinfo {author} {\bibfnamefont {K.~P.}\ \bibnamefont
  {Marinova}},\ }\href {https://doi.org/10.1016/j.adt.2011.12.006} {\bibfield
  {journal} {\bibinfo  {journal} {At. Data Nucl. Data Tables}\ }\textbf
  {\bibinfo {volume} {99}},\ \bibinfo {pages} {69} (\bibinfo {year}
  {2013})}\BibitemShut {NoStop}\bibitem [{\citenamefont {Caprio}\ \emph {et~al.}(2025)\citenamefont {Caprio},
  \citenamefont {Maris},\ and\ \citenamefont
  {Fasano}}]{caprio2025:emnorm2-part1}\BibitemOpen
  \bibfield  {author} {\bibinfo {author} {\bibfnamefont {M.~A.}\ \bibnamefont
  {Caprio}}, \bibinfo {author} {\bibfnamefont {P.}~\bibnamefont {Maris}},\ and\
  \bibinfo {author} {\bibfnamefont {P.~J.}\ \bibnamefont {Fasano}},\ }\href
  {https://doi.org/10.1103/6zk6-1sy6} {\bibfield  {journal} {\bibinfo
  {journal} {Phys. Rev. C}\ }\textbf {\bibinfo {volume} {112}},\ \bibinfo
  {pages} {044318} (\bibinfo {year} {2025})}\BibitemShut {NoStop}\bibitem [{\citenamefont {Navr\'atil}\ and\ \citenamefont
  {Barrett}(1998)}]{PhysRevC.57.3119}\BibitemOpen
  \bibfield  {author} {\bibinfo {author} {\bibfnamefont {P.}~\bibnamefont
  {Navr\'atil}}\ and\ \bibinfo {author} {\bibfnamefont {B.~R.}\ \bibnamefont
  {Barrett}},\ }\href {https://doi.org/10.1103/PhysRevC.57.3119} {\bibfield
  {journal} {\bibinfo  {journal} {Phys. Rev. C}\ }\textbf {\bibinfo {volume}
  {57}},\ \bibinfo {pages} {3119} (\bibinfo {year} {1998})}\BibitemShut
  {NoStop}\bibitem [{\citenamefont {Forss\'en}\ \emph {et~al.}(2009)\citenamefont
  {Forss\'en}, \citenamefont {Caurier},\ and\ \citenamefont
  {Navr\'atil}}]{PhysRevC.79.021303}\BibitemOpen
  \bibfield  {author} {\bibinfo {author} {\bibfnamefont {C.}~\bibnamefont
  {Forss\'en}}, \bibinfo {author} {\bibfnamefont {E.}~\bibnamefont {Caurier}},\
  and\ \bibinfo {author} {\bibfnamefont {P.}~\bibnamefont {Navr\'atil}},\
  }\href {https://doi.org/10.1103/PhysRevC.79.021303} {\bibfield  {journal}
  {\bibinfo  {journal} {Phys. Rev. C}\ }\textbf {\bibinfo {volume} {79}},\
  \bibinfo {pages} {021303} (\bibinfo {year} {2009})}\BibitemShut {NoStop}\bibitem [{\citenamefont {Caprio}\ \emph
  {et~al.}(2022{\natexlab{b}})\citenamefont {Caprio}, \citenamefont {Fasano},\
  and\ \citenamefont {Maris}}]{caprio2022:emnorm}\BibitemOpen
  \bibfield  {author} {\bibinfo {author} {\bibfnamefont {M.~A.}\ \bibnamefont
  {Caprio}}, \bibinfo {author} {\bibfnamefont {P.~J.}\ \bibnamefont {Fasano}},\
  and\ \bibinfo {author} {\bibfnamefont {P.}~\bibnamefont {Maris}},\ }\href
  {https://doi.org/10.1103/PhysRevC.105.L061302} {\bibfield  {journal}
  {\bibinfo  {journal} {Phys. Rev. C}\ }\textbf {\bibinfo {volume} {105}},\
  \bibinfo {pages} {L061302} (\bibinfo {year}
  {2022}{\natexlab{b}})}\BibitemShut {NoStop}\bibitem [{\citenamefont {N\"ortersh\"auser}\ \emph {et~al.}(2011)\citenamefont
  {N\"ortersh\"auser}, \citenamefont {Neff}, \citenamefont {S\'anchez},\ and\
  \citenamefont {Sick}}]{PhysRevC.84.024307}\BibitemOpen
  \bibfield  {author} {\bibinfo {author} {\bibfnamefont {W.}~\bibnamefont
  {N\"ortersh\"auser}}, \bibinfo {author} {\bibfnamefont {T.}~\bibnamefont
  {Neff}}, \bibinfo {author} {\bibfnamefont {R.}~\bibnamefont {S\'anchez}},\
  and\ \bibinfo {author} {\bibfnamefont {I.}~\bibnamefont {Sick}},\ }\href
  {https://doi.org/10.1103/PhysRevC.84.024307} {\bibfield  {journal} {\bibinfo
  {journal} {Phys. Rev. C}\ }\textbf {\bibinfo {volume} {84}},\ \bibinfo
  {pages} {024307} (\bibinfo {year} {2011})}\BibitemShut {NoStop}\bibitem [{\citenamefont {Esbensen}\ and\ \citenamefont
  {Bertsch}(1992)}]{esbensen1992:11li-soft-dipole}\BibitemOpen
  \bibfield  {author} {\bibinfo {author} {\bibfnamefont {H.}~\bibnamefont
  {Esbensen}}\ and\ \bibinfo {author} {\bibfnamefont {G.~F.}\ \bibnamefont
  {Bertsch}},\ }\href {https://doi.org/10.1016/0375-9474(92)90219-A} {\bibfield
   {journal} {\bibinfo  {journal} {Nucl. Phys. A}\ }\textbf {\bibinfo {volume}
  {542}},\ \bibinfo {pages} {310} (\bibinfo {year} {1992})}\BibitemShut
  {NoStop}\bibitem [{\citenamefont {Zhukov}\ \emph {et~al.}(1993)\citenamefont {Zhukov},
  \citenamefont {Danilin}, \citenamefont {Fedorov}, \citenamefont {Bang},
  \citenamefont {Thompson},\ and\ \citenamefont
  {Vaagen}}]{zhukov1993:6he-11li-halo-ground-state-properties}\BibitemOpen
  \bibfield  {author} {\bibinfo {author} {\bibfnamefont {M.~V.}\ \bibnamefont
  {Zhukov}}, \bibinfo {author} {\bibfnamefont {B.~V.}\ \bibnamefont {Danilin}},
  \bibinfo {author} {\bibfnamefont {D.~V.}\ \bibnamefont {Fedorov}}, \bibinfo
  {author} {\bibfnamefont {J.~M.}\ \bibnamefont {Bang}}, \bibinfo {author}
  {\bibfnamefont {I.~J.}\ \bibnamefont {Thompson}},\ and\ \bibinfo {author}
  {\bibfnamefont {J.~S.}\ \bibnamefont {Vaagen}},\ }\href
  {https://doi.org/10.1016/0370-1573(93)90141-Y} {\bibfield  {journal}
  {\bibinfo  {journal} {Phys. Rep.}\ }\textbf {\bibinfo {volume} {231}},\
  \bibinfo {pages} {151} (\bibinfo {year} {1993})}\BibitemShut {NoStop}\bibitem [{\citenamefont {Bertsch}\ \emph {et~al.}(1990)\citenamefont
  {Bertsch}, \citenamefont {Esbensen},\ and\ \citenamefont
  {Sustich}}]{PhysRevC.42.758}\BibitemOpen
  \bibfield  {author} {\bibinfo {author} {\bibfnamefont {G.}~\bibnamefont
  {Bertsch}}, \bibinfo {author} {\bibfnamefont {H.}~\bibnamefont {Esbensen}},\
  and\ \bibinfo {author} {\bibfnamefont {A.}~\bibnamefont {Sustich}},\ }\href
  {https://doi.org/10.1103/PhysRevC.42.758} {\bibfield  {journal} {\bibinfo
  {journal} {Phys. Rev. C}\ }\textbf {\bibinfo {volume} {42}},\ \bibinfo
  {pages} {758} (\bibinfo {year} {1990})}\BibitemShut {NoStop}\bibitem [{\citenamefont {Dasgupta}\ \emph {et~al.}(1994)\citenamefont
  {Dasgupta}, \citenamefont {Mazumdar},\ and\ \citenamefont
  {Bhasin}}]{PhysRevC.50.R550}\BibitemOpen
  \bibfield  {author} {\bibinfo {author} {\bibfnamefont {S.}~\bibnamefont
  {Dasgupta}}, \bibinfo {author} {\bibfnamefont {I.}~\bibnamefont {Mazumdar}},\
  and\ \bibinfo {author} {\bibfnamefont {V.~S.}\ \bibnamefont {Bhasin}},\
  }\href {https://doi.org/10.1103/PhysRevC.50.R550} {\bibfield  {journal}
  {\bibinfo  {journal} {Phys. Rev. C}\ }\textbf {\bibinfo {volume} {50}},\
  \bibinfo {pages} {R550} (\bibinfo {year} {1994})}\BibitemShut {NoStop}\bibitem [{\citenamefont {Ueta}\ \emph {et~al.}(1999)\citenamefont {Ueta},
  \citenamefont {Miyake},\ and\ \citenamefont {Bund}}]{PhysRevC.59.1806}\BibitemOpen
  \bibfield  {author} {\bibinfo {author} {\bibfnamefont {K.}~\bibnamefont
  {Ueta}}, \bibinfo {author} {\bibfnamefont {H.}~\bibnamefont {Miyake}},\ and\
  \bibinfo {author} {\bibfnamefont {G.~W.}\ \bibnamefont {Bund}},\ }\href
  {https://doi.org/10.1103/PhysRevC.59.1806} {\bibfield  {journal} {\bibinfo
  {journal} {Phys. Rev. C}\ }\textbf {\bibinfo {volume} {59}},\ \bibinfo
  {pages} {1806} (\bibinfo {year} {1999})}\BibitemShut {NoStop}\bibitem [{\citenamefont {Kumar}\ and\ \citenamefont
  {Bhasin}(2002)}]{PhysRevC.65.034007}\BibitemOpen
  \bibfield  {author} {\bibinfo {author} {\bibfnamefont {S.}~\bibnamefont
  {Kumar}}\ and\ \bibinfo {author} {\bibfnamefont {V.~S.}\ \bibnamefont
  {Bhasin}},\ }\href {https://doi.org/10.1103/PhysRevC.65.034007} {\bibfield
  {journal} {\bibinfo  {journal} {Phys. Rev. C}\ }\textbf {\bibinfo {volume}
  {65}},\ \bibinfo {pages} {034007} (\bibinfo {year} {2002})}\BibitemShut
  {NoStop}\bibitem [{\citenamefont {Brida}\ \emph {et~al.}(2006)\citenamefont {Brida},
  \citenamefont {Nunes},\ and\ \citenamefont {Brown}}]{brida2006effects}\BibitemOpen
  \bibfield  {author} {\bibinfo {author} {\bibfnamefont {I.}~\bibnamefont
  {Brida}}, \bibinfo {author} {\bibfnamefont {F.~M.}\ \bibnamefont {Nunes}},\
  and\ \bibinfo {author} {\bibfnamefont {B.~A.}\ \bibnamefont {Brown}},\ }\href
  {https://doi.org/10.1016/j.nuclphysa.2006.06.012} {\bibfield  {journal}
  {\bibinfo  {journal} {Nucl. Phys. A}\ }\textbf {\bibinfo {volume} {775}},\
  \bibinfo {pages} {23} (\bibinfo {year} {2006})}\BibitemShut {NoStop}\bibitem [{\citenamefont {Id~Betan}(2017)}]{betan2017cooper}\BibitemOpen
  \bibfield  {author} {\bibinfo {author} {\bibfnamefont {R.~M.}\ \bibnamefont
  {Id~Betan}},\ }\href {https://doi.org/10.1016/j.nuclphysa.2017.01.004}
  {\bibfield  {journal} {\bibinfo  {journal} {Nucl. Phys. A}\ }\textbf
  {\bibinfo {volume} {959}},\ \bibinfo {pages} {147} (\bibinfo {year}
  {2017})}\BibitemShut {NoStop}\bibitem [{\citenamefont {Garrido}\ and\ \citenamefont
  {Jensen}(2020)}]{PhysRevC.101.034003}\BibitemOpen
  \bibfield  {author} {\bibinfo {author} {\bibfnamefont {E.}~\bibnamefont
  {Garrido}}\ and\ \bibinfo {author} {\bibfnamefont {A.~S.}\ \bibnamefont
  {Jensen}},\ }\href {https://doi.org/10.1103/PhysRevC.101.034003} {\bibfield
  {journal} {\bibinfo  {journal} {Phys. Rev. C}\ }\textbf {\bibinfo {volume}
  {101}},\ \bibinfo {pages} {034003} (\bibinfo {year} {2020})}\BibitemShut
  {NoStop}\bibitem [{\citenamefont {Garrido}\ \emph {et~al.}(2023)\citenamefont
  {Garrido}, \citenamefont {Jensen}, \citenamefont {Fynbo},\ and\ \citenamefont
  {Riisager}}]{PhysRevC.107.014003}\BibitemOpen
  \bibfield  {author} {\bibinfo {author} {\bibfnamefont {E.}~\bibnamefont
  {Garrido}}, \bibinfo {author} {\bibfnamefont {A.~S.}\ \bibnamefont {Jensen}},
  \bibinfo {author} {\bibfnamefont {H.~O.~U.}\ \bibnamefont {Fynbo}},\ and\
  \bibinfo {author} {\bibfnamefont {K.}~\bibnamefont {Riisager}},\ }\href
  {https://doi.org/10.1103/PhysRevC.107.014003} {\bibfield  {journal} {\bibinfo
   {journal} {Phys. Rev. C}\ }\textbf {\bibinfo {volume} {107}},\ \bibinfo
  {pages} {014003} (\bibinfo {year} {2023})}\BibitemShut {NoStop}\bibitem [{\citenamefont {Casten}(2000)}]{casten2000:ns}\BibitemOpen
  \bibfield  {author} {\bibinfo {author} {\bibfnamefont {R.~F.}\ \bibnamefont
  {Casten}},\ }\href
  {https://doi.org/10.1093/acprof:oso/9780198507246.001.0001} {\emph {\bibinfo
  {title} {Nuclear Structure from a Simple Perspective}}},\ \bibinfo {edition}
  {2nd}\ ed.,\ \bibinfo {series} {Oxford Studies in Nuclear Physics}\
  No.~\bibinfo {number} {23}\ (\bibinfo  {publisher} {Oxford University
  Press},\ \bibinfo {address} {Oxford},\ \bibinfo {year} {2000})\BibitemShut
  {NoStop}\bibitem [{\citenamefont {Suhonen}(2007)}]{suhonen2007:nucleons-nucleus}\BibitemOpen
  \bibfield  {author} {\bibinfo {author} {\bibfnamefont {J.}~\bibnamefont
  {Suhonen}},\ }\href {https://doi.org/10.1007/978-3-540-48861-3} {\emph
  {\bibinfo {title} {From Nucleons to Nucleus}}}\ (\bibinfo  {publisher}
  {Springer-Verlag},\ \bibinfo {address} {Berlin},\ \bibinfo {year}
  {2007})\BibitemShut {NoStop}\bibitem [{\citenamefont {Rowe}\ and\ \citenamefont
  {Wood}(2010)}]{rowe2010:rowanwood}\BibitemOpen
  \bibfield  {author} {\bibinfo {author} {\bibfnamefont {D.~J.}\ \bibnamefont
  {Rowe}}\ and\ \bibinfo {author} {\bibfnamefont {J.~L.}\ \bibnamefont
  {Wood}},\ }\href {https://doi.org/10.1142/6209} {\emph {\bibinfo {title}
  {Fundamentals of Nuclear Models: Foundational Models}}}\ (\bibinfo
  {publisher} {World Scientific},\ \bibinfo {address} {Singapore},\ \bibinfo
  {year} {2010})\BibitemShut {NoStop}\bibitem [{\citenamefont {Maris}\ \emph {et~al.}(2015)\citenamefont {Maris},
  \citenamefont {Caprio},\ and\ \citenamefont {Vary}}]{maris2015:berotor2}\BibitemOpen
  \bibfield  {author} {\bibinfo {author} {\bibfnamefont {P.}~\bibnamefont
  {Maris}}, \bibinfo {author} {\bibfnamefont {M.~A.}\ \bibnamefont {Caprio}},\
  and\ \bibinfo {author} {\bibfnamefont {J.~P.}\ \bibnamefont {Vary}},\ }\href
  {https://doi.org/10.1103/PhysRevC.91.014310} {\bibfield  {journal} {\bibinfo
  {journal} {Phys. Rev. C}\ }\textbf {\bibinfo {volume} {91}},\ \bibinfo
  {pages} {014310} (\bibinfo {year} {2015})}\BibitemShut {NoStop}\bibitem [{\citenamefont {Zuker}\ \emph {et~al.}(1995)\citenamefont {Zuker},
  \citenamefont {Retamosa}, \citenamefont {Poves},\ and\ \citenamefont
  {Caurier}}]{zuker1995:shell-rotation-quasi-su3}\BibitemOpen
  \bibfield  {author} {\bibinfo {author} {\bibfnamefont {A.~P.}\ \bibnamefont
  {Zuker}}, \bibinfo {author} {\bibfnamefont {J.}~\bibnamefont {Retamosa}},
  \bibinfo {author} {\bibfnamefont {A.}~\bibnamefont {Poves}},\ and\ \bibinfo
  {author} {\bibfnamefont {E.}~\bibnamefont {Caurier}},\ }\href
  {https://doi.org/10.1103/PhysRevC.52.R1741} {\bibfield  {journal} {\bibinfo
  {journal} {Phys. Rev. C}\ }\textbf {\bibinfo {volume} {52}},\ \bibinfo
  {pages} {R1741} (\bibinfo {year} {1995})}\BibitemShut {NoStop}\bibitem [{\citenamefont {Navr\'{a}til}\ \emph {et~al.}(2026)\citenamefont
  {Navr\'{a}til}, \citenamefont {Quaglioni}, \citenamefont {Hupin},
  \citenamefont {Gennari},\ and\ \citenamefont
  {Kravvaris}}]{navratil2026:halo-ab-initio-halo25}\BibitemOpen
  \bibfield  {author} {\bibinfo {author} {\bibfnamefont {P.}~\bibnamefont
  {Navr\'{a}til}}, \bibinfo {author} {\bibfnamefont {S.}~\bibnamefont
  {Quaglioni}}, \bibinfo {author} {\bibfnamefont {G.}~\bibnamefont {Hupin}},
  \bibinfo {author} {\bibfnamefont {M.}~\bibnamefont {Gennari}},\ and\ \bibinfo
  {author} {\bibfnamefont {K.}~\bibnamefont {Kravvaris}},\ }\href
  {https://doi.org/10.3390/particles9020057} {\bibfield  {journal} {\bibinfo
  {journal} {Particles}\ }\textbf {\bibinfo {volume} {9}},\ \bibinfo {pages}
  {57} (\bibinfo {year} {2026})}\BibitemShut {NoStop}\end{thebibliography}
\end{document}